\documentclass{emulateapj}
\newcommand{\Mb}{M_\mathrm{B}}
\newcommand{\mb}{m_\mathrm{B}}
\slugcomment{Accepted by The Astronomical Journal}
\shorttitle{Virgo early-type dwarfs. II. Blue centers}
\shortauthors{Lisker, Glatt, Westera, \& Grebel}
\begin{document}
 
\title{Virgo cluster early-type dwarf galaxies with the Sloan Digital
  Sky Survey.\\
II. Early-type dwarfs with central star formation}

\author{Thorsten Lisker, Katharina Glatt, Pieter Westera, and Eva
  K. Grebel}
\affil{Astronomical Institute, Dept.\ of Physics and Astronomy,
  University of Basel, Venusstrasse 7, CH-4102 Binningen, Switzerland}
\email{tlisker@astro.unibas.ch}
 
\begin{abstract}
Despite the common picture of an early-type dwarf (dE) as a quiescent galaxy
with no star formation and little gas, we identify 23 dEs that have blue central
colors caused by recent or ongoing star formation in our sample of 476
Virgo cluster dEs. In addition, 14 objects that were mostly classified as
(candidate) BCDs have similar properties.
 Among the certain cluster members, the dEs with blue centers reach a
fraction of more than 15\% of the dE population at brighter ($\mb\le16$)
magnitudes. A spectral analysis of the centers of 16 galaxies reveals in
all cases an underlying old population that dominates the mass, with
$M_{\rm old}\ge90\%$ for all but one object. Therefore the majority of these
galaxies will appear like ordinary dEs within $\sim$ one Gigayear or less
after the last episode of star formation. Their overall gas content is less
than that of dwarf irregular galaxies, but higher than that of ordinary
dEs. Their flattening distribution suggests the shape of a thick disk,
similar to what has 
been found for dEs with disk features in Paper I of this series. Their
projected spatial distribution shows no central clustering, and
  their distribution with projected local density follows that of irregular
  galaxies, indicative of an unrelaxed population. This is
corroborated by their velocity distribution, which displays two side
peaks characteristic of recent infall.
 We
discuss possible formation mechanisms (ram-pressure
stripping, tidally induced star formation, harassment) that might be
able to explain both the disk shape and the central star formation of
the dEs with blue centers.
\end{abstract}
 
\keywords{
 galaxies: dwarf ---
 galaxies: structure ---
 galaxies: evolution ---
 galaxies: stellar content ---
 galaxies: clusters: individual (Virgo) ---
 galaxies: fundamental parameters
}
 
%________________________________________________________________

\section{Introduction}
 \label{sec:intro}

Early-type dwarf galaxies are the
most numerous type of galaxy in clusters, suggesting that the vigorous
forces acting within a cluster environment might actually be creating
them. At least theoretically, this could come about via a morphological
transformation of infalling galaxies, like e.g.\ 
ram-pressure stripping of irregular galaxies \citep{gun72}, or
so-called harassment of late-type spirals \citep{moo96}. While
these scenarios would in principle also be able to explain
the famous morphology-density relation \citep[e.g.][]{dre80},
unambiguous observational proofs are difficult to obtain. Early-type
dwarfs are characterized by their smooth, regular appearance,
while any possible progenitor galaxy probably displays a different
overall structure, before a morphological transformation occurs.
Moreover, early-type dwarfs themselves are not a homogeneous class of
objects. In addition to the classical dwarf ellipticals,
\citet{san84} introduced the class of dwarf S0 (dS0)
galaxies, which were conjectured to have
disk components, based on indications like high flattening or a
bulge-disk-like profile \citep{bin91}.
Following up on the discovery of spiral structure in an
early-type dwarf \citep{jer00a} and on similar other discoveries, we
identified 41 Virgo cluster early-type dwarfs with disk features in
\citet[][hereafter Paper I]{p1}, and argued that they constitute a
population of disk-shaped objects. Furthermore,
nucleated and non-nucleated dwarf ellipticals show 
significantly different clustering properties \citep{bin87} and 
flattening distributions \citep{ryd94,bin95}.

Aside from this structural heterogeneity, color differences have been
reported as well \citep{rak04,lis05}, hinting at a range of stellar
populations in early-type dwarfs. While the classical dwarf ellipticals
are typically considered not to have recent or ongoing star formation
and no significant gas or dust content \citep[e.g.][]{gre01,conIV},
the well known Local Group galaxy NGC 205 constitutes a prominent example
for an early-type dwarf that does exhibit central star formation, gas,
and dust
\citep{hod73}. More examples are NGC 185 \citep{hod63} and the
apparently isolated galaxy IC 225 \citep{gu06}. Furthermore,
\citet{conIV} report a 15\% HI detection rate for early-type dwarfs in
the Virgo cluster. All these objects are morphologically early-type
dwarfs, but form a heterogeneous family of objects, which we
assign the common abbreviation ``dE'', thereby including galaxies
classified as dwarf elliptical or as dS0.

Based on the general idea of gas removal due to effects
like ram-pressure stripping, numerous observational studies have
focused on comparing the properties of gas-rich dwarf irregular (dIrr) galaxies
and gas-poor dEs and dwarf
spheroidals (dSphs), attempting to confirm or reject a possible evolutionary
relation between them. \citet{thu85} found in his study of
optical--near-infrared colors the metallicity ranges of dIrrs and
Virgo dEs to be ``mutually 
exclusive''.
In their spectroscopic analysis of oxygen abundances of planetary nebulae
  in Local Group dEs and dSphs, \citet{ric98} also found a significant
  offset to the respective abundances of dIrrs.
 Similarly, \citet{gre03} recently showed for the Local
Group that this metallicity difference exists even when considering
only the respective old stellar populations of dIrrs and
dSphs,
indicative of more intense star formation and enrichment in dSphs.
\citet{thu85} and \citet{bin85} also noted that many Virgo dEs
are nucleated, while none of 
the dIrrs are.
\citet{bot86} 
concluded that Virgo dIrrs would have to fade by $\sim$1.5 mag in B to
fall into the optical--near-infrared color range of dEs. However, the
bulk of faded dIrrs 
would then have an effective surface brightness $\mu_{\rm B,e} >
25 $mag/arcsec$^2$, substantially lower than the observed values of
dEs.

Despite these counter-arguments, potential dIrr/dE ``transition
types'' have frequently been discussed \citep[e.g.][]{fer94,joh97,kne99}.
Recently, \citet{vZe04a} pointed out the ``remarkable commonality''
between dEs and dIrrs with respect to their surface brightness distributions
and metallicity-luminosity relations. These authors found significant
rotation in several bright dEs, upon which they base their discussion about
the possibility that these very dEs may have formed via ram-pressure
stripping of dIrrs.
While \citet{vZe04a} admitted that dEs (in Virgo) might actually form
through various processes, they argued that
every dE must have been gas-rich and star-forming in the past, and
would thus inevitably have been classified as dIrr.
However, these properties also apply to blue compact dwarf (BCD)
  galaxies, which have often been discussed in the
literature as potential dE progenitors.

An important difference between dIrrs and BCDs was highlighted by
\citet{bot86}, who characterized dIrrs as an ``odd combination of rather
blue colors, yet quite low surface brightness,'' indicative of a low
surface mass density. In contrast, BCDs have a much higher surface
brightness, suggesting a higher surface mass density, since they
are not bluer than the dIrrs. Whether or not BCDs could indeed be the
gas-rich, star-forming progenitors of dEs has been discussed
controversially \citep[e.g.][]{bot86,dri96,guz96,pap96}.
A glance at the color images\footnote{Using the Sloan Digital Sky Survey Image List
Tool, http://cas.sdss.org/astro/en/tools/chart/list.asp, Authors: J.\
Gray, A.\ Szalay, M.\ Nieto-Santisteban, and T.\ Budavari}
 of Virgo cluster galaxies that were
classified as (candidate) BCDs reveals what might be one
reason for this controversy: these objects do not constitute a
homogeneous class, but they vary strongly in size, (ir)regularity, and
the fraction of area dominated by blue light with respect to the
total area of the galaxy. It might thus be more promising to simply look for
plausible dE progenitors among the BCDs, instead of attempting to draw
conclusions about this class as a whole.

Interestingly, \citet{bot86} mentioned that the BCDs in their sample
look similar to NGC 205. Could dEs with central star formation thus
bridge the evolutionary gap from quiescent dEs to potentially
star-bursting progenitors? 
To our knowledge,
\citet{vig84} were the first to identify a central star formation
region in a dwarf elliptical outside the Local Group. Recently,
\citet{gu06} presented a similar dE with ongoing star formation in its
center. In Paper I of this series \citep{p1}, we identified nine
Virgo early-type dwarfs with central irregularities likely to be caused
by dust or gas, and found that all of them have a blue center. In the present
paper (Paper II), we follow up on these objects, and systematically
search the Virgo cluster for such dEs with blue
centers. This is made possible by the publicly available data of the
Sloan Digital Sky Survey (SDSS) Data Release 4 \citep[DR4,][]{sdssdr4}
which covers almost the whole Virgo cluster with multiband optical
imaging and partly with spectroscopy. Our data and sample are
  described in Sects.~\ref{sec:data} and ~\ref{sec:sample}, respectively.
Results from image analysis are presented in
Sect.~\ref{sec:images}, followed by the spectral analysis in
Sect.~\ref{sec:spectra}. Section~\ref{sec:gas} focuses on
the gas content of our galaxies. The
systematic properties of dEs with blue centers are given
in Sect.~\ref{sec:stuff}. Their evolutionary role is discussed
in Sect.~\ref{sec:discuss}, followed by a
summary and outlook in Sect.~\ref{sec:summary}.

%________________________________________________________________

\section{Data}
 \label{sec:data}

\subsection{SDSS imaging}
\label{sec:sub_sdssima}

  The SDSS DR4 covers
  all galaxies listed in the Virgo Cluster Catalog \citep[VCC, ][]{vcc}
  with 
  a declination of $\delta \lesssim 16\fdg25$, except for an
  approximately $2\arcdeg \times 2\fdg5$ area at $\alpha\approx 186\fdg2$,
  $\delta\approx +5\fdg0$. It provides reduced
  and calibrated images taken in the 
  u, g, r, i, and z bands with an effective exposure time of
  $54 \rm{s}$ in each band \citep[see also][]{sdssedr}. The pixel scale
  of $0\farcs396$ 
  corresponds to a physical size of $30 \rm{pc}$ when adopting a Virgo
  cluster distance of $d=15.85 \rm{Mpc}$, i.e.\ a distance modulus
  $m-M=31.0$ mag \citep[see e.g.][]{fer00}, which we use throughout. The
  SDSS imaging camera   \citep{sdsscamera}
  takes data in drift-scanning 
  mode nearly simultaneously in the five photometric bands, and thus
  combines very homogeneous multicolor photometry 
  with large area coverage, good resolution, and sufficient depth to
  enable a systematic analysis of early-type dwarfs.
  The images have an absolute astrometric accuracy of ${\rm RMS} \le
  0\farcs1$ per coordinate, and a relative accuracy between the r band
  and each of the other bands of less than $0.1$ pixels
  \citep{sdssastrometry}. They can thus easily be aligned using
  their astrometric calibration and need not be registered manually.
  The RMS of the noise per pixel corresponds to a
  surface brightness of approximately $24.2 {\rm mag/arcsec^2}$ in the
  u-band, $24.7$ in g, $24.4$ in r, $23.9$ in i, and $22.4$ in z.
  The typical total signal-to-noise ratio 
  (S/N) of a bright dE ($\mb\approx 14$) amounts to about $1000$
  in the r-band within an aperture radius of
  approximately two half-light radii. For a faint dE ($\mb \approx 18$) this
  value is typically about $50$. While the S/N in the g and i-band is
  similar to the above value, it is several times lower in the z-band
  and more than ten times lower in the u-band. Therefore
  u and z will not be used in the following analysis.

The SDSS provides photometric measurements for our galaxies, but
we found these to be incorrect in many cases \citep{lis05}. The SDSS
photometric pipeline significantly overestimates the local sky flux
around the Virgo dEs due to their large apparent sizes and low surface
brightness outskirts. This affects the measurement of isophotal and
Petrosian radii, the profile fits, and subsequently the calculation of
total magnitudes, which can be wrong by up to 0.5 mag.
Therefore, we use $B$ magnitudes from the VCC throughout the
paper, and defer calculation of total magnitudes from SDSS data to a
more detailed photometric study of the Virgo dEs.

\subsection{SDSS spectroscopy}
\label{sec:sub_sdssspec}

 The centers of several galaxies are also covered by the SDSS
 spectroscopic survey, which provides fiber spectra with a wavelength
 coverage of $3800 - 9200$ \AA\
 and a resolution of 1800 or larger. The spectra are binned in
 logarithmic wavelength, such that the wavelength interval of one
 pixel along the dispersion axis corresponds to a constant velocity
 interval of 69 km s$^{-1}$ \citep{sdsstechn}.
 The fiber diameter corresponds to an angular size of
 $3\arcsec$, which translates into $231 $pc at a distance
 $d=15.85 \rm{Mpc}$. Typical half-light
 radii of the brighter dEs range from $\sim 10\arcsec$ to $\sim
 25\arcsec$ \citep{bin91}, or from 0.8 to 1.9 kpc. \emph{Therefore,
 spectral information is only available for the very central region
 of the dEs}.

\subsection{Radial velocities}
\label{sec:sub_velodata}

Heliocentric velocities are available for 198 dEs of our sample of 414
dEs from Paper I that are listed as certain Virgo cluster
members in \citet{vcc} and \citet{virgokin}.
Velocities were
taken from the NASA/IPAC Extragalactic Database (NED),
originally provided by the SDSS and by the following studies:
\citet{vcc,1991trcb.book.....D,1992ApJS...83...29S,virgokin,1995MNRAS.273.1141Y,dri96,1998ApJS..119..277G,1999PASP..111..438F,2000A&AS..146..259G,2000A&AS..144..463V,conI,2002A&A...384..371S,2003AJ....125.2891C,geh03,2004A&A...417..499G}.

Four of these galaxies, however, have velocities above 6000 km s$^{-1}$
 from recent data (SDSS and \citealt{conI}). Since
 the velocities of known Virgo cluster members are lower by more than a
 factor of two, we change the membership status of these galaxies (VCC
 0401, VCC 0838, VCC 1111, VCC 1517) to ``possible member''. This
 leaves us with a subsample of 194 dEs that are certain cluster members
 and for which radial velocity measurements are available.

\section{Sample}
\label{sec:sample}

\subsection{Sample selection}
\label{sec:sub_sample}

In Paper I of this series, we put together a sample of Virgo cluster
dEs that comprises all dEs with 
$\mb\le18.0$ mag that are listed in the VCC, that are covered by the SDSS, and
that are certain or possible cluster members according to \citet{vcc} and
\citet{virgokin}. While objects with uncertainties were initially included,
we then excluded all galaxies that appeared to be possible dwarf irregulars due to
asymmetric shapes. Objects classified as ``dE/dIrr'' were
not included. 25 galaxies classified as (candidate) dE in
the VCC are not covered by the SDSS DR4. The resulting sample comprises
476 early-type dwarfs, 410 of which are certain cluster members (see
Sect.~\ref{sec:sub_velodata}). Note that our magnitude limit of
$\mb\le18.0$ mag corresponds to the magnitude up to which the VCC was
found to be complete \citep{vcc}. With a distance modulus of $m-M=31.0$
mag, this translates into $\Mb\le-13.0$ mag.

In this sample, we identified several dEs with irregular or
clumpy central features likely caused by gas and dust (cf.\ Fig.~6 of
Paper I). By dividing the background-subtracted aligned g
and i-band images provided by Paper I, we obtain color maps that are not
calibrated, yet are useful to look for significant color gradients. 
Most of the objects just described have a central region whose color
is clearly bluer than that of the rest of the galaxy, similar to the dE
recently presented by \citet{gu06}. Using these
color maps, we visually searched all 476 early-type dwarf
galaxies that were presented in Paper I for such a blue center.
As a complementary check we then
examined the radial $g-i$ color profiles (Sect.~\ref{sec:sub_imatechn})
of the thus selected galaxies to confirm the presence of a clear color
gradient.

23 out of 476 dEs (16/410 certain cluster
members) entered our working sample of galaxies with blue centers
(Figs.~\ref{fig:pics1a}, 
\ref{fig:pics1b}, and \ref{fig:pics1c}).
We shall term these objects ``dE(bc)s'' hereafter.
Note that the presence of \emph{weak} color gradients (either negative or
positive) in dEs has been reported in the literature
\citep[e.g.][]{gav05}. Such objects are not the focus of this
study; we rather aim at dEs with a blue center, i.e. a significant
positive gradient.
The distinction between a weak and
a significant gradient might appear somewhat arbitrary;
however, a detailed and quantitative study of color gradients of our full dE sample is
beyond the scope of this paper and will be presented elsewhere. The
weakest gradient visually selected by us is that of VCC 0308,
with a color difference of 0.1 mag between inner and outer regions
 (see
Fig.~\ref{fig:pics1b}).

We point out that the dE(bc)s were morphologically classified as dwarf
ellipticals or dS0s by \citet{san84}, and were confirmed as early-type
dwarfs in our Paper I. While \citet{fer06} suggest to reclassify four
dE(bc)s as dE/dIrr based on their blue central colors and irregular
isophotal shapes in the center, we do not use color as a criterion for
\emph{morphological} classification. The \emph{central} irregularities
are the reason for which they were assigned to the class dS0
\citep{bin91}, but their overall appearance is smooth and regular, as
can be seen from the combined images (Figs.~\ref{fig:pics1a} to
\ref{fig:pics1c}). Complementary to these, we show isophotal contours
of the dE(bc)s in Fig.~\ref{fig:isomain} (see
Sect.~\ref{sec:sub_imatechn}). There, the central 
irregularities in several dE(bc)s can be seen, which are also revealed 
by the unsharp mask images. Outside of
the central 
region, though, the isophotes are regularly shaped, confirming the
early-type dwarf morphology of these galaxies.

Our preselection of dEs relies on the classification
given in the VCC, and on the subsequent
  examination of the combined images in Paper I.
However, the intial selection was based on
photographic plates, which are most sensitive to the blue light.
Therefore, if the color distribution of a candidate dE(bc) would be quite
asymmetric or if the blue central region would make up for a rather
large fraction of the total light, this galaxy might not have been 
classified as dE in the first place.\footnote{Note that the dE(bc)s identified above
    were classified as dE \emph{despite} their having been observed
    in blue light, in which the light of young
    stars -- if present -- dominates.}
We thus decided to use the color-combined images provided by the
SDSS Image List
Tool\footnote{http://cas.sdss.org/astro/en/tools/chart/list.asp, Authors: J.\
Gray, A.\ Szalay, M.\ Nieto-Santisteban, and T.\ Budavari}
to search all objects classified as irregular
galaxy or as blue compact dwarf (BCD; including candidates) for
galaxies that look similar to the dE(bc)s, i.e., that have a regular
outer shape and a color similar to the dEs, while having a blue inner
region. We found 12 such objects (10
certain cluster members). Furthermore, two
galaxies classified as E/S0 with magnitudes
similar to the brighter dEs show the same appearance; both are certain
cluster members. Images and color profiles of these additional
galaxies are presented in Figs.~\ref{fig:pics2a} and \ref{fig:pics2b},
and their isophotal contours are shown in Fig.~\ref{fig:isoadd}.

 To ensure that we are
not mixing different types of galaxies, we shall use separate samples of
objects in the course 
of our study: the \emph{main sample} comprising the 23
galaxies classified as dE that have a blue center, and an
\emph{additional sample}
comprising the 14 galaxies classified other than dE that are
  similar in appearance to the dE(bc)s.
Table~\ref{tab:bludE} lists our selected objects along with
their classification.

\subsection{Presence of disks}
\label{sec:sub_dEdi}

In Paper I we presented a systematic search for disk features in Virgo
cluster dEs, which we detected in 41 out of 476 objects (including
candidates). We showed that these galaxies -- termed dE(di)s -- are not
simply dwarf ellipticals that just have an embedded disk component, but
appear to be instead a population of genuine disk galaxies, i.e.\ flat
oblate objects. Amongst the
23 dE(bc)s of our main sample, five are (candidate) dE(di)s, or 4 out of 16 if
only certain cluster members are counted. The fraction of dE(di)s among
the full dE sample is roughly about 25\% at the median $B$ magnitude of
the dE(bc)s ($14.86$, again counting only certain cluster members). In
a randomly chosen sample of 16 dEs we 
would thus expect to find 4 dE(di)s, which equals our observed number. 
Two galaxies in the additional sample also show
possible disk features (VCC 0135: possibly spiral arms, VCC 1437:
possibly spiral arms and a bar). Both are certain cluster members, so
the number of 2 out of 10 objects with disks is again consistent with
the above fractions.
 A detailed
comparison of the properties of dE(bc)s, dE(di)s, and ordinary dEs is
presented in Sect.~\ref{sec:stuff}.

%________________________________________________________________

\section{Image analysis}
 \label{sec:images}

\subsection{Techniques}
\label{sec:sub_imatechn}

The study presented in Paper I provides us with
background-subtracted, aligned g, r, and i-band images, a combined
image, unsharp masks, and an elliptical aperture
for each object. A detailed description can be found in
Paper I. Briefly, combined images were obtained by co-adding the g,
r, and i-band images to increase the S/N. From these we produced
unsharp masks with various kernel sizes
(using \emph{IRAF}\footnote{IRAF is distributed by
  the National Optical Astronomy Observatories, which are operated by
  the Association of Universities for Research in Astronomy, Inc.,
  under cooperative agreement with the National Science
  Foundation.}\emph{/gauss};
  \citealt{iraf}), as well as isophotal contour
diagrams (using \emph{IRAF/newcont}).
An elliptical aperture for
each galaxy was determined by performing ellipse fits with
\emph{IRAF\,/\,ellipse} on the combined image, and then choosing by
eye one of the outer elliptical isophotes to trace best the outer shape
of the galaxy. This isophote was usually between 1 and 2 half-light
radii.

The SDSS flux calibration was applied to the aligned g and i-band
images for each source, following the instructions on the SDSS
webpage\footnote{http://www.sdss.org/dr4/algorithms/fluxcal.html}.
These images were then divided by each other, converted into
magnitudes, and corrected for Galactic extinction \citep{sch98},
yielding proper $g-i$ color maps.
From the same images we obtained radial intensity profiles by
azimuthally averaging over elliptical annuli with
\emph{IRAF/ellipse}. We used geometric steps (i.e., steps that increase
by a constant factor) and a fixed position angle, 
ellipticity, center, and semi-major axis. The ratio of the g
and i-band fluxes was then converted into magnitudes, yielding radial
$g-i$ color profiles, where radius is calculated from semi-major ($a$)
and semi-minor axis ($b$) as $\sqrt{ab}$. Disturbing
foreground stars and background galaxies were masked prior to profile
calculation.

\subsection{Results from the image analysis}
\label{sec:sub_imaresults}

\begin{figure}
  \epsscale{0.9}
  \plotone{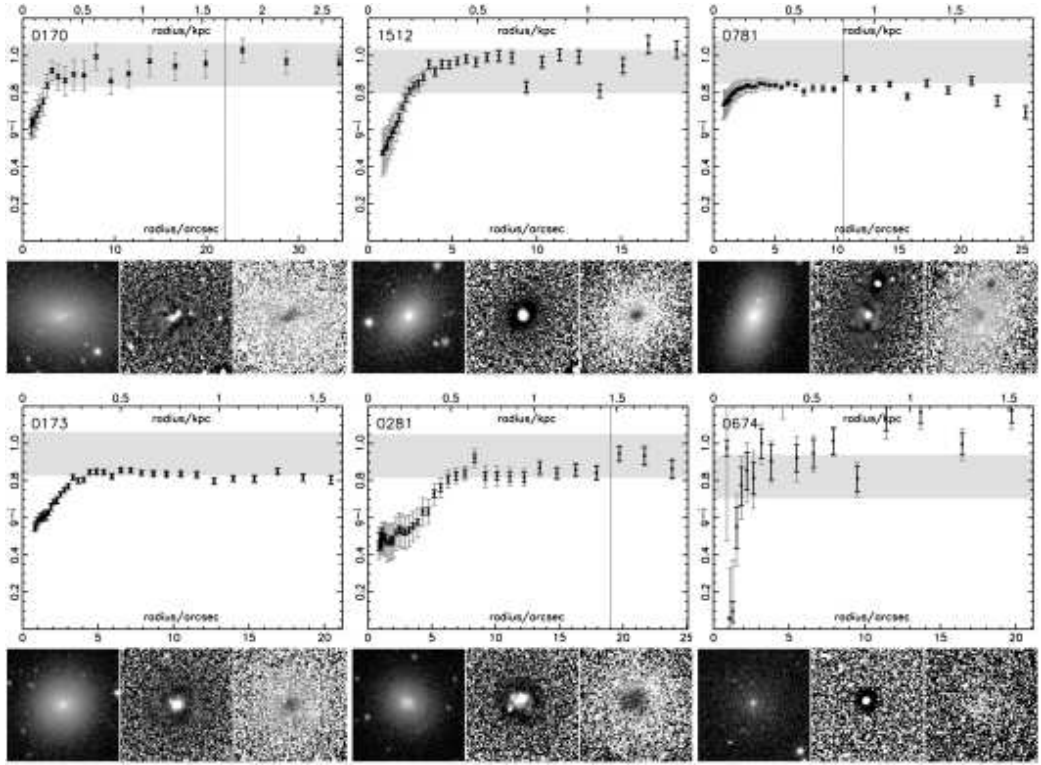}
  %\begin{center}
  %\includegraphics[width=14cm]{f1.eps}
  %\end{center}
  \caption{
Early-type dwarfs with blue centers.
      For each galaxy, the radial $g-i$ color profile is shown (top
      panel; radius$=\sqrt{ab}$), along with the combined image (bottom left), the unsharp
      mask created with a Gaussian filter with $\sigma=4$ pixels (bottom center), and
      the $g-i$ color map (bottom right).
      %177 pixels bzw. 88.5 pixels
      %->70 arcsecs bzw. 35 arcsecs
      %->5.4kpc bzw. 2.7kpc 
      The combined images have a horizontal scale of $70''$, or $5.4\,$kpc with
    $d=15.85\,\rm{Mpc}$, while the scales of unsharp masks and color maps are
      only half as large. A legend showing the grey scales of the color maps
      is given in Fig.~\ref{fig:pics1c}.
   The galaxies are sorted such that those with a relatively constant
      outer color and an abruptly starting, steep gradient come first,
      while those with a gradual color change come last (in
      Fig.~\ref{fig:pics1c}). The sorting 
      has been done visually, without any quantitative basis.
      The black error bars in the color profiles give
the uncertainty calculated from the S/N only, whereas the grey error
bars represent the \emph{azimuthal} variation of the color at the
respective radius. Since the latter includes S/N-effects, the grey error
      bars are always larger than the black ones. The
      vertical dotted line denotes the half-light radius -- if
      available -- as given by \citet{bin93}. The profiles are shown up to
      the estimated radius from \citet{vcc},
      at $\mu_{\rm B} \approx 25.5 $mag/arcsec$^2$.
      The grey-shaded areas enclose the 2-$\sigma$-range of
      the colors of ``ordinary'' dEs (i.e.\ without a blue center) at the
      respective magnitude, as derived from \citet{lis05}. See text for
      details. For VCC 0674, the steps between each point are twice as
      large as for the other galaxies, due to its low S/N.
    }
\label{fig:pics1a}
\end{figure}

\begin{figure}
  \epsscale{0.9}
  \plotone{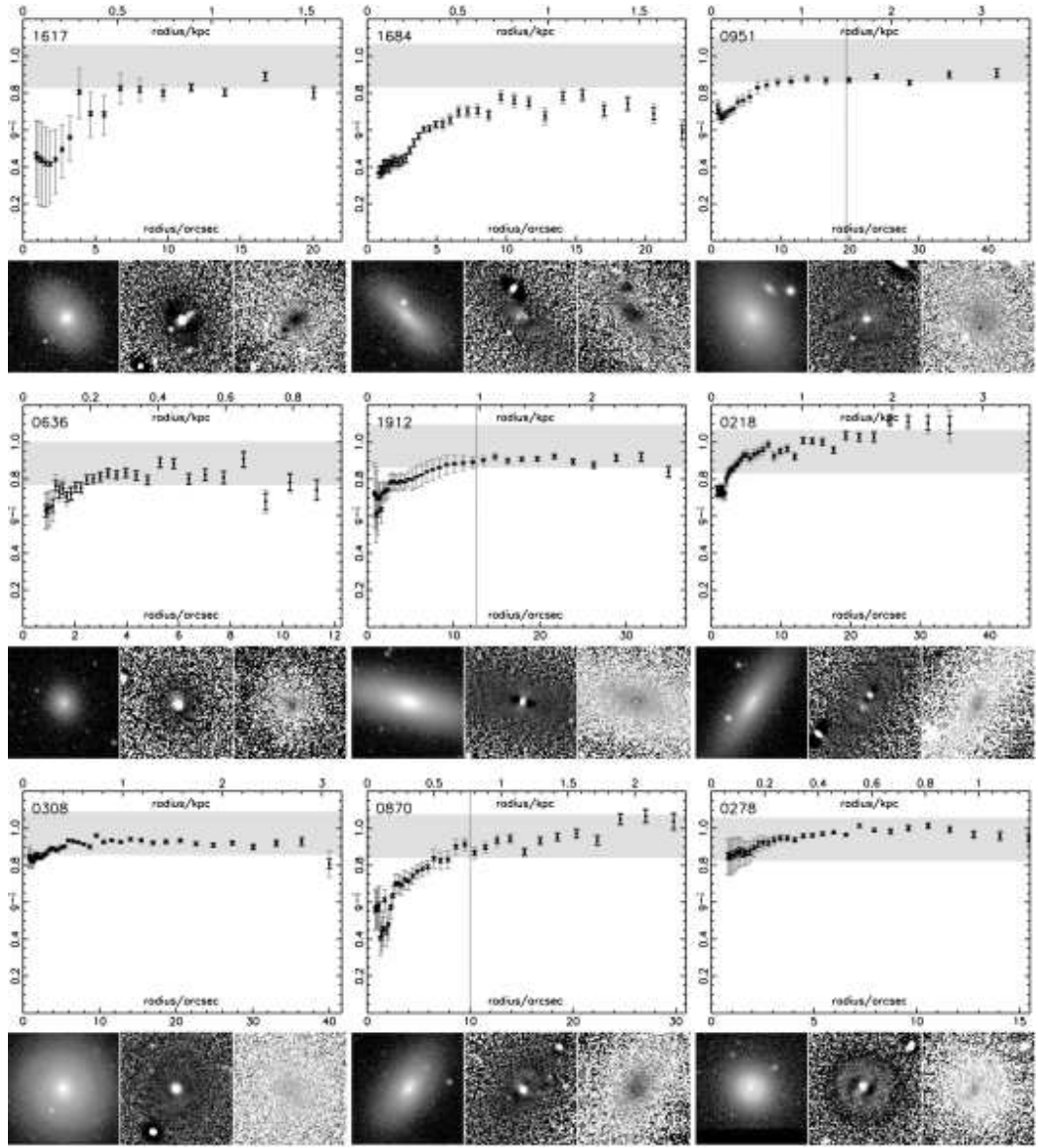}
  \caption{Early-type dwarfs with blue centers.
Continued from Fig.~\ref{fig:pics1a}.
    }
\label{fig:pics1b}
\end{figure}

\begin{figure}
  \epsscale{0.9}
  \plotone{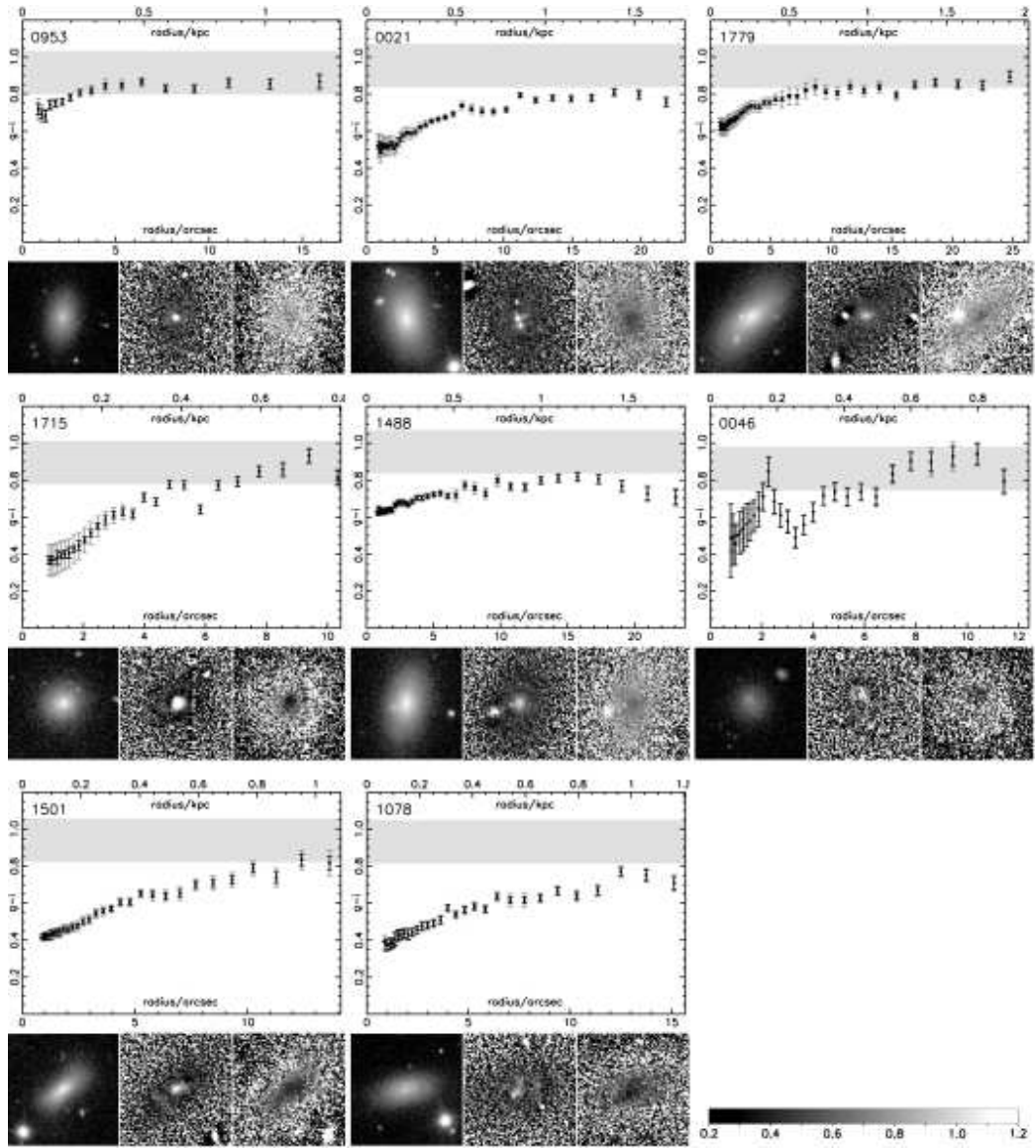}
  \caption{Early-type dwarfs with blue centers.
Continued from Fig.~\ref{fig:pics1b}. The bar in the bottom right
    panel indicates the grey scale used for the color maps.
    }
\label{fig:pics1c}
\end{figure}

\begin{figure}
  \epsscale{0.9}
  \plotone{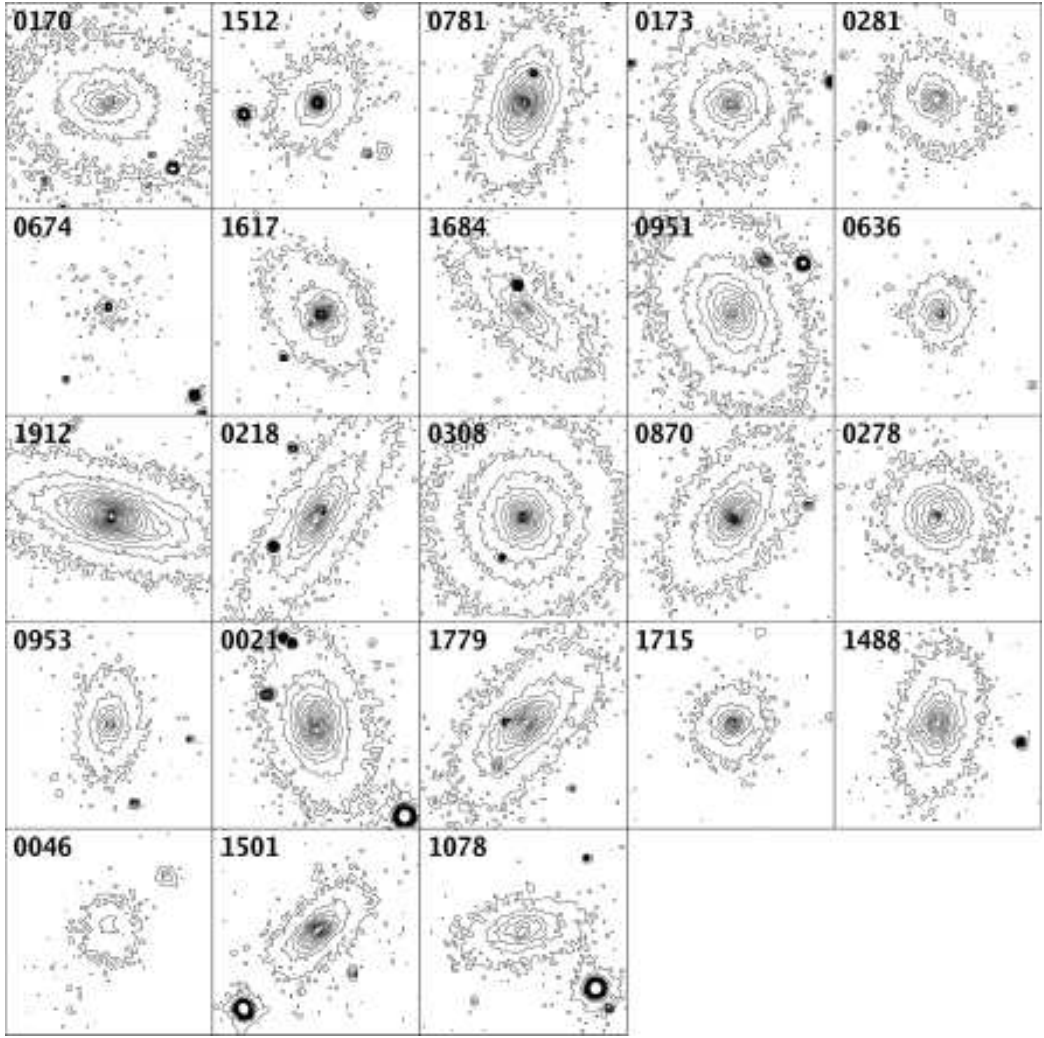}
  \caption{Isophotal contours of the dE(bc)s.
      The galaxies are shown in the same order as in Figs.~\ref{fig:pics1a}
      to \ref{fig:pics1c}. Contour diagrams were produced with
      \emph{IRAF\,/\,newcont} on the combined images. The outermost contour
      lies at a level of three times the noise RMS, which was measured for
      each image separately. Contours are then displayed for 15 logarithmic
      steps up to 300 times the RMS.
  }
  \label{fig:isomain}
\end{figure}

\begin{figure}
  \epsscale{0.9}
  \plotone{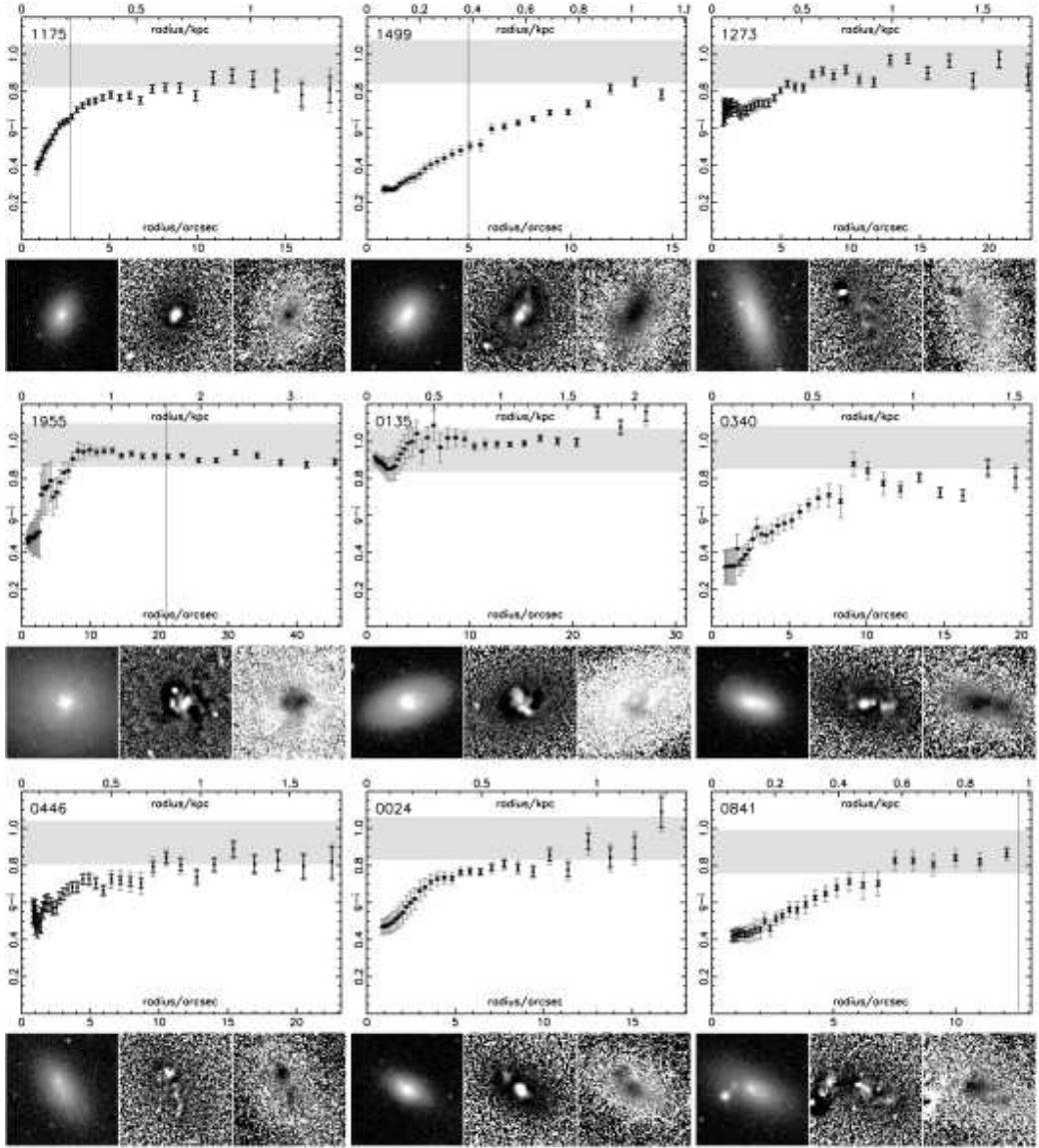}
  \caption{Galaxies similar to the dE(bc)s.
Like Fig.~\ref{fig:pics1a}, but for the galaxies of the additional
sample, which were not
classified as dEs but were chosen by us as being similar. The
two top left galaxies were classified as E or S0; the top right galaxy
as irregular. All other objects here and in Fig.~\ref{fig:pics2b} were
classified as (candidate) BCDs, and are sorted according to the shape
of the color gradient, analogous to Figs.~\ref{fig:pics1a} to
\ref{fig:pics1c}.
    }
\label{fig:pics2a}
\end{figure}

\begin{figure}
  \epsscale{0.9}
  \plotone{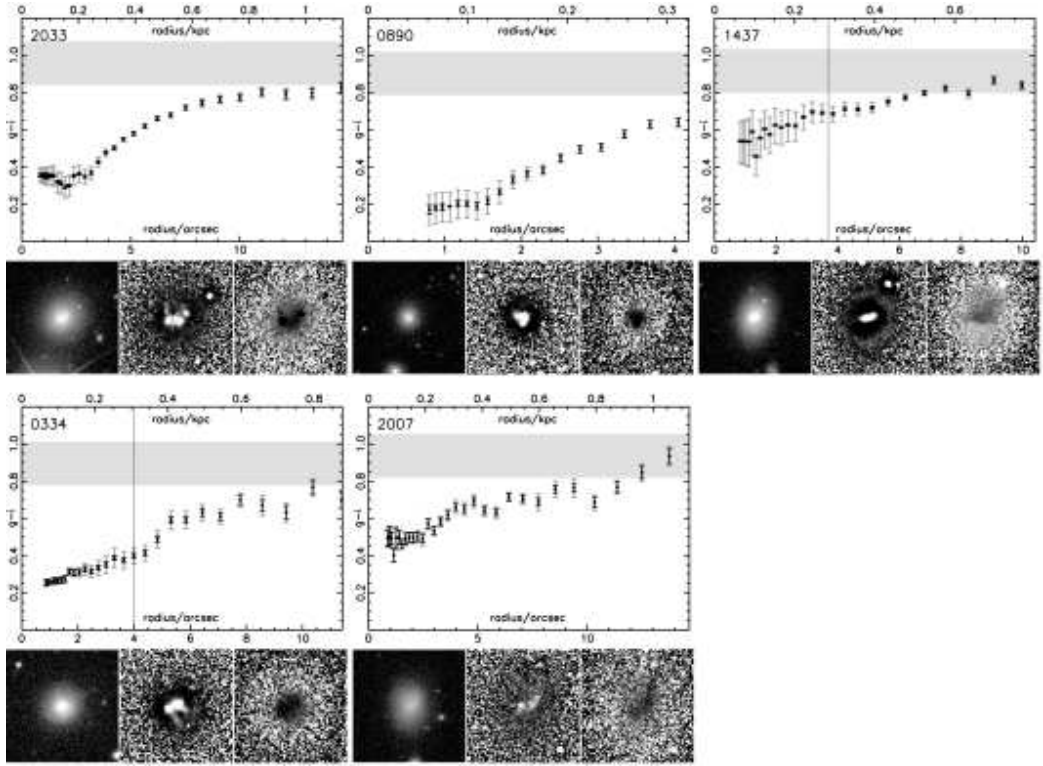}
  \caption{Galaxies similar to the dE(bc)s.
Continued from Fig.~\ref{fig:pics2a}.
    }
\label{fig:pics2b}
\end{figure}

\begin{figure}
  \epsscale{0.9}
  \plotone{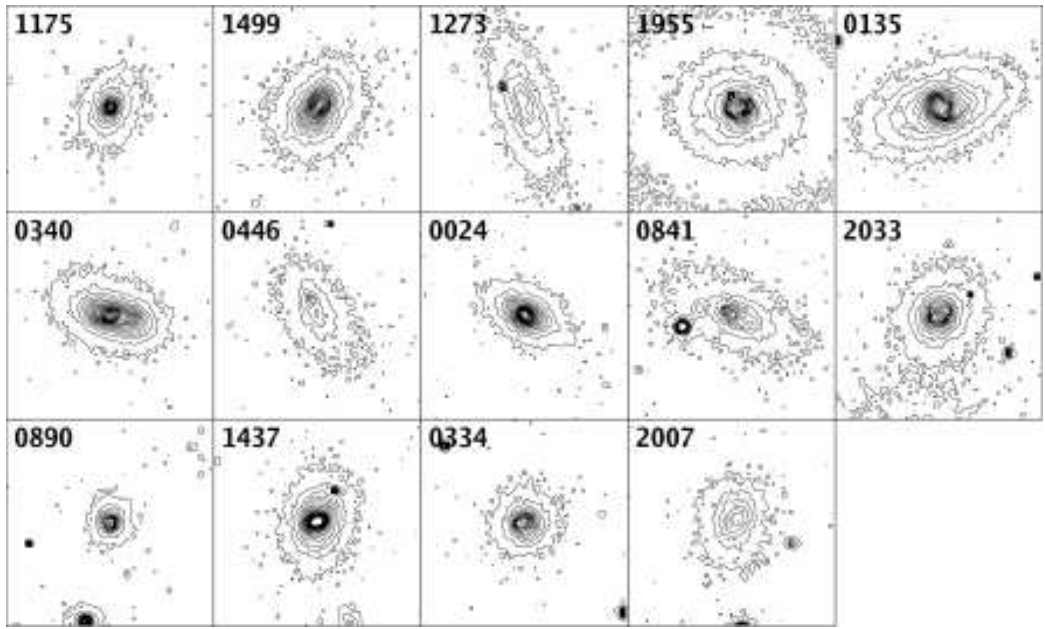}
  \caption{Isophotal contours of the additional sample.
      Like Fig.~\ref{fig:isomain}, but for the galaxies of the
      additional sample, which are shown in the same order as in
      Figs.~\ref{fig:pics2a} and \ref{fig:pics2b}.
  }
  \label{fig:isoadd}
\end{figure}

In Figs.~\ref{fig:pics1a} to \ref{fig:pics2b} we present for each galaxy
the radial color profile, the combined image, the unsharp mask created
with a Gaussian filter with $\sigma=4$ pixels, and the color
map. The
color profiles contain two types of errors: the black error bars give
the uncertainty calculated from the S/N only, whereas the grey error
bars represent the \emph{azimuthal} variation of the color at the
respective radius (which, of course, includes S/N-effects). A color
distribution that is not symmetric with respect to the 
galaxy center (e.g.\ in VCC 1617, see Fig.~\ref{fig:pics1b}) or is
somewhat irregular (e.g.\ in VCC 0170, see Fig.~\ref{fig:pics1a}) leads
to an azimuthal variation of color significantly larger 
than the uncertainty from S/N only.

In many cases the unsharp masks reveal central irregularities likely
caused by gas and dust features or by asymmetric star forming
regions. On average, the central irregularities are stronger and the
blue central regions are larger for galaxies of the additional sample
as compared to those of main sample.

Obviously, some galaxies have a relatively constant outer color and abruptly
start to become bluer when going inwards (e.g.\ VCC 0173), while
others display a gradual color change from the outer regions to the
center (e.g.\ VCC 1501). Although it is not always
unambiguous which of the two cases applies to an object, we attempted
to sort the dE(bc)s in Figs.~\ref{fig:pics1a} to \ref{fig:pics1c}, as
well as the (candidate) BCDs of the additional sample in
Figs.~\ref{fig:pics2a} and \ref{fig:pics2b}, 
according to the color 
profile shape: starting with those that have a constant outer color until a
steep gradient sets in abruptly, to those with a smooth gradient. We
found no correlations of this profile shape with either magnitude, radius,
or surface brightness.

In order to compare the $g-i$ colors of the dE(bc)s with those of
``ordinary'' dEs, we make use of the colors computed by \citet{lis05}
for 228 galaxies of our full dE sample, excluding dE(bc)s. Those colors
were derived from aperture photometry on the SDSS images, using
circular apertures with a radius equalling the half-light radius. We
performed a linear fit to 
the color-magnitude relation of $B$ magnitude versus $g-i$ color,
yielding an average dE color for each magnitude
with a standard deviation of 0.06 mag.
These
values are shown in Figs.~\ref{fig:pics1a} to \ref{fig:pics2b} for each
galaxy: 
the grey shaded bands enclose the 2-$\sigma$ range of
color at each galaxy's magnitude.
For most 
galaxies, the outer color is still bluer than
the typical dE color; for several objects it is even bluer than the
2-$\sigma$ range. Note that the above fit has been performed on dE
colors computed \emph{within} the half-light radius; however, no
strong gradient is to be expected since all dE(bc)s have been excluded
from the fit.
Thus, the relatively blue outer colors of the dE(bc)s could hint at a younger age of
the dE(bc)s as a whole, a lower metallicity, or a shorter time since
the last star formation activity in the outer regions as compared to
ordinary dEs. A 
spectroscopic examination of the stellar content of the dE(bc)s is
possible at least for the centers of several galaxies, as presented in
the following section.

%________________________________________________________________

\section{Spectral analysis}
 \label{sec:spectra}

In order to explore the stellar content of our
objects, we examine integrated spectra from the SDSS DR4 as
described in Sect.~\ref{sec:sub_sdssspec}. These are taken with fibers
of a diameter of $3\arcsec$, corresponding to a physical size of 231 pc
at a distance $d=15.85 \rm{Mpc}$.
10 such spectra are available for the main dE(bc) sample, one of which
(VCC 0046) proved too noisy for a stellar population analysis. Seven spectra are
available for the additional sample. These 16 galaxies are
labelled in the last column of Table~\ref{tab:bludE} with ``S''. 
Six dE(bc)s of the main sample and six galaxies of the additional sample
display Balmer line emission (see Table~\ref{tab:bludE}). One more
dE(bc) of the main sample and three galaxies of the additional sample
are detected in the H$\alpha$ imaging study of \citet{bos02}. This is a
clear indication for ongoing star formation in the dE(bc)s. We note that
the overall emission line strengths are larger for objects of the
additional sample than for those of the main sample.

We determine the stellar content using a population synthesis method,
described in \citet{cuisinier_06}, wherein the spectra
are fitted to synthetic composite stellar populations' spectra.
Similar methods have already been employed by \citet{cidfernandes},
\citet{kong}, \citet{westera_04}, and \citet{gu06}; the latter describe
a dE with blue center very similar to our objects.

\subsection{Synthetic stellar populations}
\label{sec:sub_models}

The synthetic composite stellar populations were composed using simple
stellar population (SSP)
spectra from three different libraries of SSPs.
The first SSP library (hereafter the ``BC99'' library) was
produced using the Bruzual and Charlot 2000 Galaxy Isochrone
Spectral Synthesis Evolution Library (GISSEL) code
\citep{charlot_91,bruzual_93,bruzual_00}, implementing the Padova
2000 isochrones \citep{girardi_00} combined with the BaSeL 3.1
``Padova 2000'' stellar library \citep{paperiii,diss}.
The second SSP library, ``Starburst'',
consists of spectra from the STARBURST99 data package
\citep{leitherer} including nebular
continuum emission (Fig.~1 on the STARBURST99 web site,
http://www.stsci.edu/science/starburst99/).
It implements the 
BaSeL 2.2 library \citep{lejeune_97,lejeune_98}, and for stars with
strong mass loss it also takes 
into account extended model atmospheres by \citet{schmutz},
combined with the Geneva isochrones
\citep{meynet,schaller,schaerera,schaererb,charbonnel}.
For old populations, the ``BC99'' spectrum was used, since the Starburst99
data package only contains spectra up to 900 Myr.
Additionally to these two libraries,
% - which were already used in \citet{westera_04} -
we also used a library with higher spectral
resolution, ``BC03'', produced by employing the 2003 version of the
GISSEL code \citep{bruzual_03} and the Padova 1995
isochrones \citep{fagotto2,girardi}
combined with the STELIB
\citep{leborgne} stellar library.
The nebular continuum emission was also added to the spectra in the ``BC99''
and ``BC03'' libraries, in the same way as described by
\citet{leitherer}.
%Table~\ref{tab:libraries} provides a summary of evolutionary tracks and
%stellar libraries used for the three SSP libraries.
%

\subsection{Fit to model spectra: procedure}
\label{sec:sub_specfit}

The spectra cover a wavelength range of around 3820~\AA\ to 9200~\AA, but the
wavelength region above 8570~\AA\ is too much affected by telluric lines,
so we fit the full spectra from 3820~\AA\ to 8570~\AA, except for those
various parts of the spectra showing ``contamination'' from
different emission and/or telluric line sources.
Table~\ref{tab:cutouts} lists all the regions that were omitted.

\begin{figure}
  \epsscale{0.35}
  \plotone{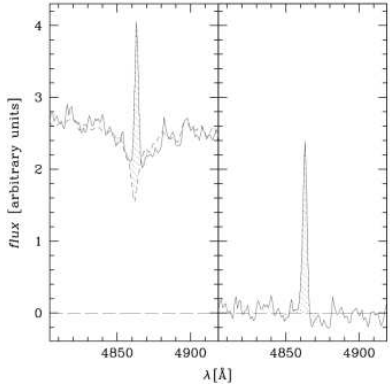}
  \caption{Emission line strength. Illustration of how the
       H$\beta$ emission line strength was 
       measured for the example of VCC 0870.
       In the left panel, the solid line represents the observed spectrum,
       whereas the short-dashed line shows the best fit using the
       ``BC03'' SSP library.
       The shaded region between these two lines shows the area used to
       calculate the emission line strength.
       The right panel shows the emission line after subtracting the
       (rebinned) best fit. The shaded area corresponds to the shaded
       area between spectrum and best fit from the left panel.}
  \label{fig:EW}
\end{figure}

We corrected the spectra for redshift and for Galactic foreground extinction
using the values from \citet{sch98} and the extinction law of
\citet{fluks}. In the eight cases with H$\alpha$ and H$\beta$ emission
we also corrected for internal gas extinction, again using the Fluks et
al.\ extinction law. 
The extinction constants $E(B-V)=A_V/3.266$ were estimated from the
$H\alpha /H\beta$
Balmer decrements following \citet{torrespeimbert},
adopting intrinsic ratios of the emission line fluxes
$I_{H\alpha}/I_{H\beta}=2.87$ \citep{osterbrock}.
In order to properly determine the emission line strengths, we first
had to remove the contribution from the absorption lines of the underlying 
stellar populations.
This was done by making a first fit to the spectra employing the method
described in the following paragraph, using the highest resolution library (``BC03'',
see above), and then subtracting the best fit spectra from the
observed spectra as illustrated in Fig.~\ref{fig:EW}. The so found
values for $E(B-V)$ were multiplied by a 
factor of 0.44 to correct for systematic differential extinction
between the stellar populations and the gas \citep{calzetti}.
In the cases where the spectrum displays only H$\alpha$ but no H$\beta$
emission, we assume the internal extinction to be negligible: in
all of these spectra the emission is much weaker than in the cases with
both H$\alpha$ and H$\beta$ emission,
in which the lowest value for $E(B-V)$ is already close to
zero ($E(B-V)=0.02$). The average value for internal extinction of the
eight galaxies with H$\alpha$ and H$\beta$ emission is
$E(B-V)=0.10$. Note that the problem of apparent 
truncation of strong emission lines in SDSS spectra reported by
\citet{kni04} does not occur in any of our spectra.

%%%%% \item Fit with 3 populations (like in Westera et al. and Gu et al.)
%
We modelled the actual
population as being composed of an old ($\ge$ 1 Gyr), an
intermediate-age (10 Myr to 1 Gyr), and a young ($<$ 10 Myr)
stellar population. 
 While there is no standard definition for this age terminology, these age
  ranges are chosen to reflect the significant changes in the spectrum
  of an SSP with increasing age. Similar values were used e.g.\ by
  \citet{cidfernandes,mar05,gu06}.
The characteristics and free parameters of the three populations are
summarized in Table~\ref{tab:parameters}.
 In order to confine the
parameter space to as few dimensions as possible and to focus on the
relative fractions of the populations, we fixed the age of
the old population at 5 Gyr, and the metallicity of all populations
at ${\rm [Fe/H]}=-0.3$. These values equal the best-fitting mean age
and metallicity found by \citet{geh03} in a study of Lick/IDS
absorption line indices for dEs. Moreover, the spectra of SSPs of
various ages do not differ much at ages of several Gyr and above
\citep[e.g.][]{bruzual_03}.

The best fitting population was found by a $\chi^2$ algorithm.
In order to be able to calculate the $\chi^2$ estimator, the observed
and theoretical spectra have to be on the same wavelength grid (at least in
the range used for the fit).
This was done by rebinning the observed spectra to the resolution of the
theoretical spectra using a gaussian kernel function with a full width at
half maximum (FWHM) corresponding to the resolution of the theoretical
spectra (20~\AA\ for ``BC99'' and ``Starburst'', and 1~\AA\ for ``BC03'').
The same was done with the ``signal-to-noise spectra'', i.e.\ the
S/N as a function of wavelength provided along with
the actual spectra. The $\chi^2$ estimator was calculated with
wavelength-dependent weighting, giving higher weight to the regions
with a higher S/N.
Fig.~\ref{fig:bestfit} shows an example of an observed spectrum and the
resulting composite model spectra, as well as the individual spectra of
the three assumed subpopulations.

\subsection{Fit to model spectra: results}
\label{sec:sub_specresults}

\begin{figure}
  \epsscale{0.85}
  \plotone{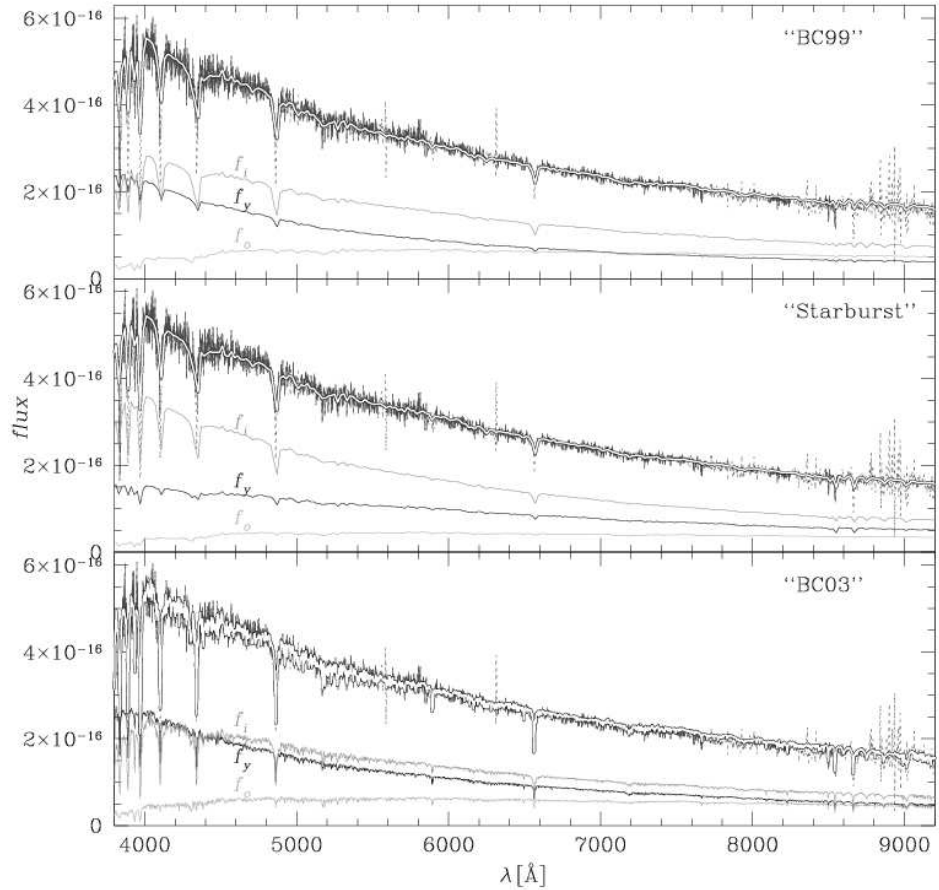}
  \caption{Example of a ``best fit''. The thin (red or dark grey) lines represent
       the observed spectrum (VCC 1499), the thicker white lines show the best
       fitting spectra obtained using the different SSP libraries, whereas
       the lower-level  lines show the best fits
       decomposed into the young ($f_y$, blue or dark grey), the
       intermediate-age ($f_i$, green or medium grey), and
       the old ($f_o$, cyan or light grey) populations. Flux is given
       in erg cm$^{-2}$ s$^{-1}$ \AA$^{-1}$.}
  \label{fig:bestfit}
\end{figure}

The best fitting population parameters for the fits with the different
libraries can be found in columns 2 to 6 of Tables~\ref{tab:EWparametersbc99}
to~\ref{tab:EWparametersbc03}.
In all dE(bc)s of the main sample, and all but one (VCC 1499) of the
additional sample, the old population makes up for 90~\% or more
of the total mass, even though its contribution to the total light is
rather small (spectra $f_o$ in Fig.~\ref{fig:bestfit}).
The young population, on the other hand, usually contributes less than
1~\% to the mass, but often dominates the light ($f_y$ in Fig.~\ref{fig:bestfit}),
whereas the intermediate-age population usually makes up for a few percent of
the stellar mass.
However, all three populations are doubtlessly present, from which we
conclude that these galaxies have been forming stars until
the present day. We remind the reader again that the spectra cover only
the very central part of our objects and do not allow us to draw direct
conclusions about the surrounding regions.

Typically, the parameter space of such spectral fits is full of
degeneracies. For example, a change of the ratio of the mass fractions of
young and intermediate-age population has a similar effect on the composite
spectrum as varying one of the ages of these populations.
Therefore, the parameter values given in
Tables~\ref{tab:EWparametersbc99} to~\ref{tab:EWparametersbc03} for
\emph{individual} galaxies should be taken
with a grain of salt.
However, the excellent agreement between the solutions found with the
different libraries, and the similarities in the solutions for the
different galaxies, suggest that the \emph{general} trends
in the derived parameters reflect the real properties of the dE(bc)s,
within the simplified framework of our three-population models.
 We emphasize that our approach of approximating the SDSS
  spectra with synthetic populations composed of three SSPs is not
  meant to represent the detailed star formation history of our target
  galaxies, but only to demonstrate the range of ages during which
  star formation must have occured. Whether stars were formed by
  short starbursts or by extended episodes of star formation can
  typically not be decided for galaxies only observable in integrated
  light \citep[e.g.][]{lil03}.

\subsection{Evolution of $g-i$ color}
\label{sec:sub_color}

Finally, we would like to know for how long after the last star formation phase
integrated spectral properties can be
distinguished from the ones of a pure old population. We focus on the
$g-i$ color since it was used for the actual selection of our objects.
For this purpose, we produced synthetic composite populations using the
three SSP libraries, and calculated the total $g-i$ color as a function of
the age of the youngest partial population as these composite populations
evolve.

\begin{figure}
  \epsscale{0.45}
  \plotone{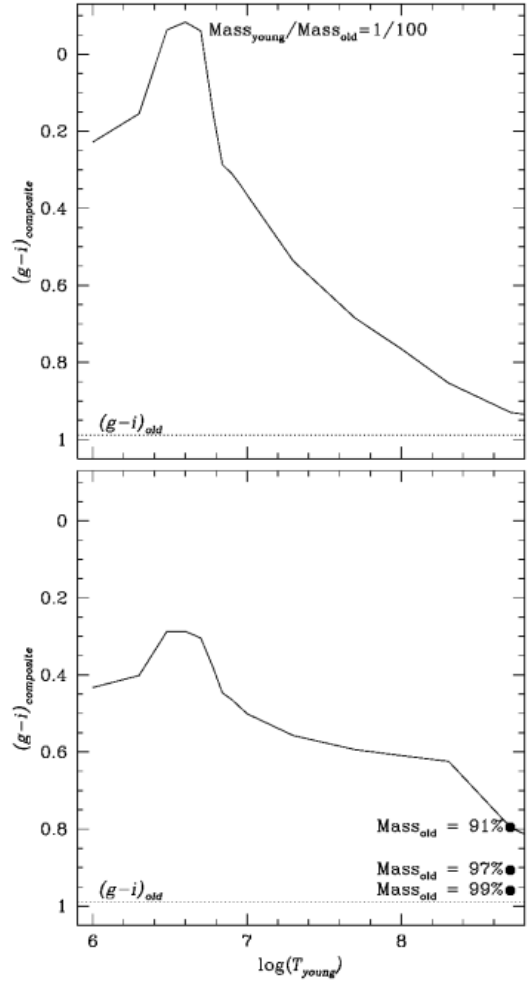}
  \caption{Color evolution of an ageing population.
      Evolution of $g-i$ color of different composite stellar
      populations. The
      age of the old population is fixed at 5 Gyr, and the metallicity
      of all populations is fixed at ${\rm [Fe/H]}=-0.3$.
      {\it Upper panel:} Composite population made out of a young and
      an old population. The young population makes up for 1\% of the
      total mass.
      {\it Lower panel:} Composite population made out of a young, an
      intermediate-age, and an old population, similar to the best fit
      results derived for the dE(bc)s. The upper curve represents a
      population with $(M_{y}+M_{i})$:$M_{o}=1$:10,
      $M_{y}$:$M_{i}=1$:30, with $M_{y,i,o}$ denoting the mass
      fractions of young, intermediate-age, and old population,
      respectively. The intermediate-age population is chosen to be $\sim$500
      Myr older than the young one. The big black dots mark the
      color of this mixture after 500 Myr, as well as that of two
      other mixtures: \{$(M_{y}+M_{i})$:$M_{o}=1$:30,
      $M_{y}$:$M_{i}=1$:30\} (middle dot), and
      \{$(M_{y}+M_{i})$:$M_{o}=1$:100, $M_{y}$:$M_{i}=1$:10\} (lower
      dot).
    }
  \label{fig:gi}
\end{figure}

The simplest case is that of two populations: a young one turning into an
intermediate-age one, on top of the old one.
If we assume the mass of the young population to have one hundredth of the mass of
the old one, the color of the mixture evolves as shown by the solid line
in the upper panel of Fig.~\ref{fig:gi}.
In this scenario, it takes a few 100 Myr until the $g-i$ color of the
composite population differs by less than 0.1 mag from a pure old
population (shown by the dotted line), and thus becomes difficult to
distinguish from the latter. The colors shown in Fig.~\ref{fig:gi} are
the ones using the ``BC99'' library, but the results hold true for all
three libraries.

Now we take instead a composite population like one of the typical solutions from
Tables~\ref{tab:EWparametersbc99} to~\ref{tab:EWparametersbc03}, that is,
$(M_{y}+M_{i})$:$M_{o}=1$:10, $M_{y}$:$M_{i}=1$:30, the intermediate-age
population being around 500 Myr older than the young one. $M_{y,i,o}$
denotes the mass fractions of young, intermediate-age, and old population,
respectively. The color of this mixture
evolves like the solid line in the lower panel of Fig.~\ref{fig:gi}.
Again, the dotted line shows the color of only an old population.
In this scenario, the mixture is still distinguishable from
a pure old population 500 Myr after the birth of the youngest
one. However, this parameter combination is the one with the smallest
mass fraction of the old population among the results for the main sample.
By using the values of the other typical solutions
from Tables~\ref{tab:EWparametersbc99} to~\ref{tab:EWparametersbc03},
$(M_{y}+M_{i})$:$M_{o}=1$:30, $M_{y}$:$M_{i}=1$:30, or
$(M_{y}+M_{i})$:$M_{o}=1$:100, $M_{y}$:$M_{i}=1$:10,
the $g-i$ color after 500 Myr is much more similar to the pure old
population color (big black dots in the lower panel of
Fig.~\ref{fig:gi}). Therefore, after the end of star formation it will
take less than $\sim$ a Gigayear for the (strong) color gradient to
disappear, and in many cases only $\sim$ half a Gigayear depending
on the mass fraction of the young and intermediate-age population. This
also demonstrates the need for a thorough study of both strong
\emph{and} weak color gradients in early-type dwarfs, in order to draw
conclusions about evolutionary histories. Whether or not star
formation will cease soon, or has ceased already in some objects,
depends on the amount (and state) of leftover gas, which is the subject of the
following section.

%________________________________________________________________

\section{Gas content}
\label{sec:gas}

Dwarf elliptical galaxies are commonly considered to be systems that
have lost their (cold) gas \citep[e.g.][]{fer94,conIV}. On the other hand,
\citet{conIV} reported ``credible'' HI detections for
seven\footnote{Two of these, however, were not classified as early-type
  dwarf in the VCC, and appear to us as at 
  least doubtful dE candidates from visual inspection of the SDSS
  images.} Virgo dEs from their own study as well as from other
literature, translating into a 15\% HI detection rate for dEs.
\clearpage
The
GOLDMine database \citep{goldmine} reports detections for six more
early-type dwarfs\footnote{Three of these 
  (VCC 0170, VCC 0227, VCC 0281) are listed as S0 or late-type spiral in
  the NASA/IPAC Extragalactic Database (NED) and similar in
  \citet{gav05} who reported their HI detections, although they were
  classified as early-type dwarf in the VCC and confirmed as such by
  our Paper I.}.
From the SDSS
spectra presented above, we know that the dE(bc)s show clear signs of
either ongoing or very recent star formation, suggesting the presence
of a certain amount of gas. It would thus be interesting
to know whether we can expect more episodes of star formation in the near
future -- requiring a significant cold gas content -- or whether star formation
is likely to cease soon due to the lack of leftover gas.

Two dE(bc)s of the
main sample are detected in HI, while for 8 others, at least upper
limits are available. Ten objects of the additional sample are
detected in HI, and upper limits are available for three further galaxies. See
Table~\ref{tab:HI} for HI masses and upper limits on 
our objects, along with the corresponding references to the literature.
Two objects have values from more than one publication; these agree well
within the errors.

We now seek to derive an estimate for the ratio of the gas mass to the total
baryonic mass, $M_{\rm HI}/M_{\rm bary}$. For this purpose we use V-band
mass-to-light ratios between 3 
and 6, as given by \citet{geh03}, and assume no dark matter content, again
following \citet{geh03}. We calculate the total V-band absolute magnitude from
the g and i-band flux within an elliptical aperture with 1.5 times the
estimated semi-major axis from \citet{vcc} (at $\sim\mu_{\rm B} =
25.5 $mag/arcsec$^2$), using the transformation of 
\citet{smi02}. The resulting gas-to-baryonic mass fractions or upper limits
for mass-to-light ratios of 3, 4.5, and 6 are listed in
Table~\ref{tab:HI}.

Of the dE(bc)s in the main sample, three objects have a very low gas content,
with $M_{\rm HI}/M_{\rm bary}$ values of 1\% or below (VCC 0170,
VCC 0951, VCC 0953). Of the remaining galaxies, one has an HI content
with a resulting fraction of $3-6$\% (depending on the adopted
mass-to-light ratio). The others have upper 
limits of several percent, up to $7-13$\% for VCC 1488.
In contrast, most of the galaxies in the additional sample have a much higher gas
fraction. Of the 10 detected galaxies, 7 reach up to more than 10\%,
with values up to $38-55$\%. However, for one object (VCC 0135) the upper limit
lies below 1\%.

A comparison with average HI-to-total mass ratios for Local Group dwarf galaxies
shows that at least some dE(bc)s of the main sample have a larger gas fraction
than the $0.2\pm 0.4$\% of an ordinary dE, while it is mostly lower than the
$30\pm 24$\%
of dwarf irregulars \citep{conIV}.
If
the detected gas was centrally concentrated, at least 
some of the dE(bc)s of the main sample might still continue to form
stars there for a significant amount of time. In contrast, most objects of the
additional sample fall in the range of the Local Group dIrrs,
suggesting that a longer duration of star formation is possible. 
 See
Sect.~\ref{sec:discuss} for a further discussion.

%________________________________________________________________

\section{Systematic properties}
\label{sec:stuff}

We have shown so far that the dE(bc)s, which were
classified morphologically as early-type dwarfs, are dominated by an
old stellar population. Given their moderate amount of cold gas and the
lack of any (significant) star formation activity beyond their central
regions, the dE(bc)s will evolve into ordinary-looking dEs
in the future. However, up to
this point we have mainly focused on the appearance and the composition
of the dE(bc)s, but we have not yet compared 
the statistical properties of the sample of dE(bc)s with those of
the sample of dEs and dE(di)s (dEs with disks, see Sect.~\ref{sec:sub_dEdi}). As
demonstrated in Paper I, the luminosity 
function, the projected spatial distribution, and the flattening
distribution are important tools to investigate differences between
types of galaxies, and to judge whether or not they constitute
separate populations. We present and analyze these distributions for
the dE(bc)s in the following subsections.

\subsection{Luminosity function}
\label{sec:sub_lum}

\begin{figure}
  \epsscale{0.45}
  \plotone{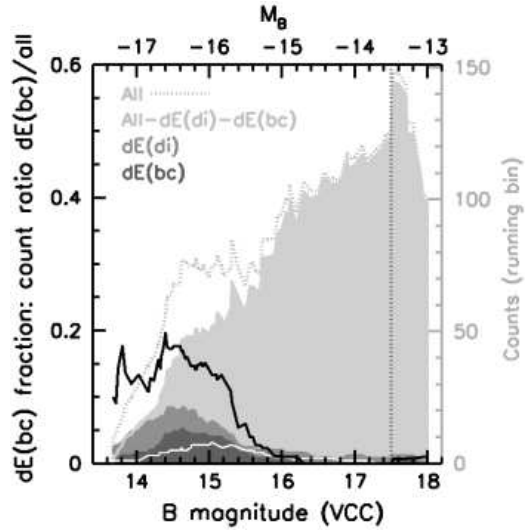}
  \caption{Luminosity function.
    Running histogram of the number of galaxies
    (right y-axis) with respect to B-band magnitude as given by the
    VCC. Shown are dE(bc)s of the main sample (dark grey area), dE(di)s
    (medium grey area), all dEs 
    excluding both dE(di)s and dE(bc)s (light grey area), and all dEs (light grey
    dashed line). Galaxies of the additional sample -- which by
    construction are not included in the dE sample -- are represented
    by the white line. The bin-width is $1\fm0$, therefore 
    the counts are incomplete for $\mb>17\fm5$ (vertical dotted
    line). A bin is calculated at each position of a galaxy in the full
    sample. Only galaxies with certain cluster membership are taken into account.
    The fraction of dE(bc)s -- the ratio of the dE(bc) histogram to
    the histogram of all dEs -- is given as black line (left y-axis
    applies).
    The upper x-axis gives absolute magnitudes assuming
    $m-M=31\fm0$.
  }
  \label{fig:maghist}
\end{figure}

   Figure~\ref{fig:maghist} shows the distribution of dE(bc)s, dE(di)s, and the
   remaining dEs with respect to their B band magnitude provided by the
   VCC. With the assumption that
   all galaxies are located at roughly the same 
   distance from us, this distribution represents their luminosity
   function. Our data
   are presented as a running histogram with a 
   bin-width of 1.0 mag (i.e.\ $\pm$0.5 mag). Only galaxies with
   certain cluster membership
   are taken into account, i.e.\ 16 dE(bc)s in the main sample whose
   distribution is given by the dark grey shaded area. The dE(di)s are represented by
   the medium grey shaded area, and the full dE sample with dE(bc)s and dE(di)s
   excluded is given by the light grey shaded area. The fraction of dE(bc)s
   among the full dE sample is shown as black line. The white line
   shows the distribution of the 12 certain cluster members of the additional
   sample, which by construction are not included in the dE sample.

   The luminosity function of the full dE sample is given as grey dashed
   line. It has a conspicuous bump at brighter magnitudes, which we explained
   in Paper I with the superposition of dE(di)s and ordinary dEs, implying that
   they were two different populations of objects. However, as can be seen in
   the figure, this bump might at least partly be explained by the superposition of
   dE(bc)s, dE(di)s, and the remaining dEs.
On the other hand, it has been shown
   in Paper I that we missed $\sim$50\% of dE(di)s due to
   issues of signal-to-noise.
   Therefore we now subtracted the
   dE(bc)s and $1.5$ times the number of dE(di)s from the full sample as a
   further test (not shown).
Still, there is no obvious over-subtraction of
   the bump (which would result in a dip), so the data are consistent
   with the estimated number of missed dE(di)s.

   The fraction of dE(bc)s among all dEs reaches up to
   more than 15\% for the main sample at brighter magnitudes, and declines to
   almost zero at $\mb \gtrsim 16$. Thus the dE(bc)s are not a
   negligible population of objects, but instead constitute a
   significant fraction of the bright dEs.
   To investigate whether the decline of the dE(bc) fraction at
   fainter magnitudes is real or whether it is due to S/N effects, we
   artificially dimmed all 11 dE(bc)s of the main sample with $14 \le \mb
   < 15$ by 1 and 2 magnitudes. This was done by adding
   Gaussian noise to the images such that the RMS of the total noise
   was increased by 1 and 2 magnitudes. Color maps and radial color
   profiles were then constructed as described above, and were examined for
   whether they would have been selected as dE(bc)s by us. When dimmed
   by 1 magnitude, 9 of the 11 objects would still have been selected,
   whereas at 2 magnitudes fainter only 5 would have been recognized
   as dE(bc)s.
   Thus, if the \emph{true} dE(bc) fraction would be constant at
   17\% (the average value within the interval $14 \le \mb < 15$)
   independent of magnitude, we would expect to \emph{observe} a
   fraction of 14\% at magnitudes between $15 \le \mb < 16$, and 8\%
   between $16 \le \mb < 17$.
  Within these magnitude ranges, we should thus find $9.4\pm2.8$
   and $8.2\pm2.8$ dE(bc)s, respectively, for a binomial
   distribution. However, we found only 4 and 0 dE(bc)s, which lies
   within 1.9 and 3.0 standard deviations, respectively. It is thus
   very likely that the decline of the number fraction of dE(bc)s is
   real.
%  - dimmed by 1 mag -> still observe 9/64 -> 14%
%    now apply these 14% to the real number of dEs within 15...16
%    (namely 67 in total) -> expect to observe 9.38 +/- 2.84
%                         -> really observed 4/67 -> within 1.9 sigma
%
%  - dimmed by 2 mag -> still observe 5/64 -> 8%
%    now apply these 8% to the real number of dEs within 16...17
%    (namely 103 in total) -> expect to observe 8.24 +/- 2.75
%                          -> really observed 0/103 -> within 3.0 sigma

   Since we are using B band
   magnitudes, it is an obvious question how much fainter the dE(bc)s
   will become after their star formation has ceased.
   From our spectral analysis, we expect a decrease of
   the B band flux
   \emph{of the central $3\arcsec$} to one third of the current flux
   (median value: 0.32) for the main sample dE(bc)s.
   To obtain this estimate we compared the total spectrum of each
   galaxy with the model spectrum of the old population only, as a
   conservative estimate. Since our galaxies show a color gradient
   (Figs.~\ref{fig:pics1a} to \ref{fig:pics1c}),
   which we interpret as a decrease of the fraction of young stars when
   going outwards, we now assume that the above ratio of faded and
   current flux linearly increases to 1 up to the B band half-light
   radius. This leads to an estimated fading of the total galaxy flux
   in the B band of 0.2 mag for all dE(bc)s. Therefore, we do not expect
   significant    evolution of the dE(bc) luminosity function after the
   cessation of    star formation.
%Zetrum: Faktor x
%Pixelanzahl geht in Ringen linear mit r: 2*Pi*r*dr
%factor(r) = 1 - (1-x) * (1 - r/hlrad)   fuer r <= hlrad, 1 sonst
%(factor so, dass z.B. 0.3 anteil alte pop.)
%

   For the galaxies in the additional sample, Figs.~\ref{fig:pics2a}
   and \ref{fig:pics2b} show that the blue regions are more
   extended. We therefore assume the linear increase of the ratio of faded and
   current flux (see above) to extend to two half-light radii. This leads
   to an estimated fading of $0.4-0.5$ mag. On average, these galaxies are already
   slightly fainter than the dE(bc)s of the main sample (see
   Fig.~\ref{fig:maghist}). This difference 
   would thus become more pronounced after their star formation
   has ceased. On the other hand, one might expect from their 
   gas content that a significant amount of stars could still be formed
   in these objects, which would counteract the fading to some extent.

\subsection{Flattening distribution}
\label{sec:sub_flat}

 The flattening distribution of the dE(bc)s is presented in
 Fig.~\ref{fig:flat}, along with that of the dE(di)s from Paper I. The
 axial ratios were derived from the elliptical apertures provided by
 Paper I (see Sect.~\ref{sec:sub_imatechn}).  Again,
 only galaxies with certain cluster membership are considered. The distributions are
   presented as running histograms with a binsize of 0.1 ($\pm 0.05$),
 normalized to an area of 1. In the bottom right hand panel of the
 figure, we show various theoretical curves for comparison. We assume 
    randomly distributed inclinations and oblate (black lines) or
 prolate (grey lines) intrinsic axial ratios. For the oblate case we
 adopt Gaussian distributions of various widths with a mean axial
 ratio of 0.4, while for the prolate case a mean value of 0.65 is
 used.

 In Paper I, we deduced that dE(di)s are consistent with being flat
 oblate objects. Here, we find
 that the dE(bc)s of the main sample are similarly distributed, with slightly
 larger axial 
 ratios than the dE(di)s. A comparison with the theoretical curves
 demonstrates that, despite the small sample size, their distribution is
 hardly consistent with dE(bc)s being spheroidal objects, and instead
 suggests the shape of a relatively thick disk. The galaxies of the
 additional sample are rounder, though not spheroidal. This is another
 point -- besides the overall gas content, 
 strength of emission lines, and size of star forming regions -- in which
 the dE(bc)s of the main sample and the additional galaxies are not alike.

\subsection{Spatial distribution}
\label{sec:sub_spatial}

\begin{figure}
  \epsscale{0.4}
  \plotone{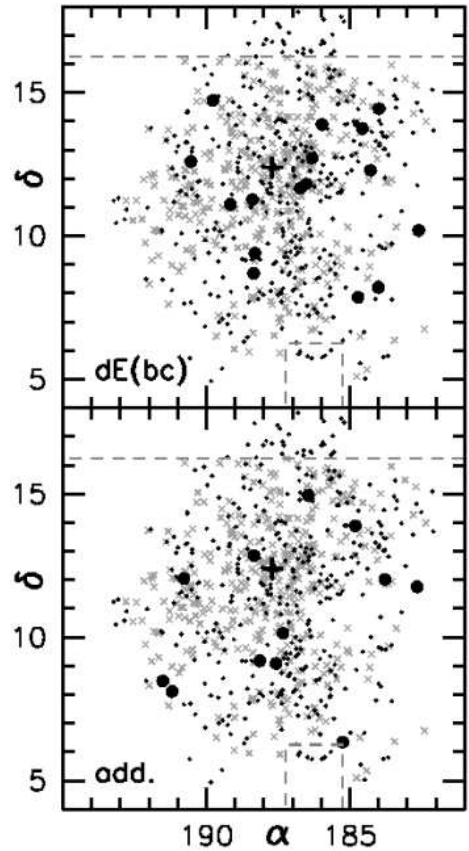}
  \caption{Distribution within the cluster.
      Projected spatial distribution of the dE(bc)s of the main sample
      (black circles, upper panel) and the galaxies of the additional sample
      (black circles, lower panel).
      Coordinates are given for J2000.     Grey crosses represent all
      dEs without blue centers. All other 
     Virgo cluster galaxies with $\mb\le18\fm0$ are shown as small
     black dots. Only galaxies with certain cluster membership are considered.
     The black cross gives the position of M87.
     Boundaries of the SDSS coverage are shown
     as grey dashed lines.
  }
  \label{fig:spatial}
\end{figure}

   In Paper I we found that while ordinary dEs are more strongly
   clustered towards the Virgo cluster center, dE(di)s basically show no
   clustering at all, another indication for them being a different
   population of galaxies.
   The dE(bc)s of the main sample also show no central clustering
   (Fig.~\ref{fig:spatial}, upper panel), similar to the dE(di)s.
   The same applies to the objects of
   the additional sample (Fig.~\ref{fig:spatial}, lower
   panel). While the number of dE(bc)s is
   relatively small and does not allow statistically secure statements,
   their 
   distribution is hardly consistent with a centrally concentrated
   population.

\subsection{Morphology-density relation}
\label{sec:sub_morphdens}

\begin{figure}
  \epsscale{0.4}
  \plotone{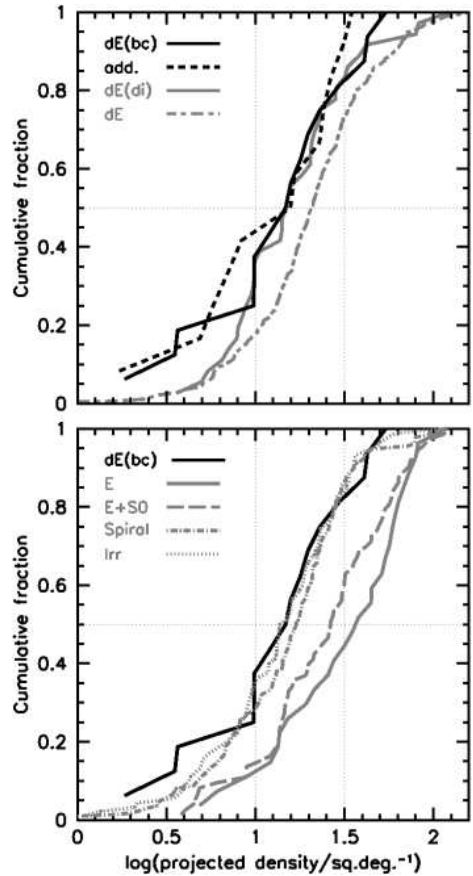}
  \caption{Morphology vs.\ density.
      Cumulative distribution of local projected densities. Following
      \citet{dre80} and \citet{bin87}, we define a circular area around
      each VCC galaxy that includes its ten nearest neighbours
      (independent of galaxy type), yielding a projected density
      (number of galaxies per square degree). Only galaxies with
      certain cluster membership are taken into account.
      \emph{Upper panel:} The dE(bc)s of the main sample (solid black
      line), the galaxies of the additional sample (dashed black line),
      the dE(di)s from Paper I (solid grey line), and ordinary dEs
      (excluding dE(bc)s and dE(di)s, dashed grey line).
      \emph{Lower panel:} The dE(bc)s of the main sample (solid black
      line) compared to various Hubble types.
  }
  \label{fig:dens}
\end{figure}

\begin{figure}
  \epsscale{0.45}
  \plotone{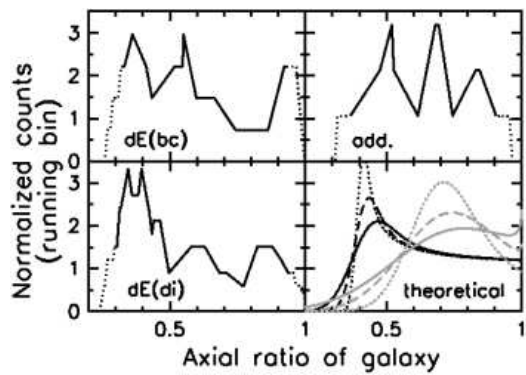}
  \caption{Flattening distribution.
    Distribution of projected axial ratios of the dE(bc)s of the main
    sample (top left), the galaxies of the additional sample
    (top right), and the dE(di)s from Paper I (bottom left).
      The data are shown as running histogram with
      a bin-width of 0.1. Only  galaxies with certain cluster
    membership are considered. Beyond the last data point on each side,
    the histograms 
    are plotted with dotted instead of solid lines. The
    bottom right hand panel shows theoretically expected curves for
    intrinsic oblate (black) and prolate (grey) shapes, assuming
    randomly distributed inclinations and intrinsic axial ratios
    that are described by the following Gaussian distributions: oblate
    with mean $\mu=0.4$ and $\sigma=0.07$ (black solid line),
    $\sigma=0.04$ (black dashed line), $\sigma=0.02$ (black dotted
    line); prolate with $\mu=0.65$ and $\sigma=0.2$ (grey solid line),
    $\sigma=0.15$ (grey dashed line), $\sigma=0.1$ (grey dotted line).
  }
  \label{fig:flat}
\end{figure}

In Fig.~\ref{fig:dens} we present the cumulative distributions of
 projected local densities -- calculated as in \citet{dre80} and
\citet{bin87} -- for the dE(bc)s, the galaxies of the additional
sample, the dE(di)s, and the remaining, ordinary dEs (upper panel). We
also compare these with the distributions of standard Hubble types
(lower panel), i.e., with the well-known morphology-density relation.

The dE(bc)s and the objects of the additional sample are
preferentially found in regions of moderate to lower density, and are
distributed similarly to the irregular galaxies.
This
implies that they, as a population, are far from being virialized,
corroborating the indications from the spatial distribution
(Sect.~\ref{sec:sub_spatial}). The dE(di)s show a similar distribution,
which follows that of the spiral galaxies. Ordinary dEs, though, are
preferentially found in
higher density regions. Their location in the morphology-density
diagram is intermediate between E/S0 galaxies and spirals.

\subsection{Velocity distribution}
\label{sec:sub_velo}

\begin{figure}
  \epsscale{0.45}
  \plotone{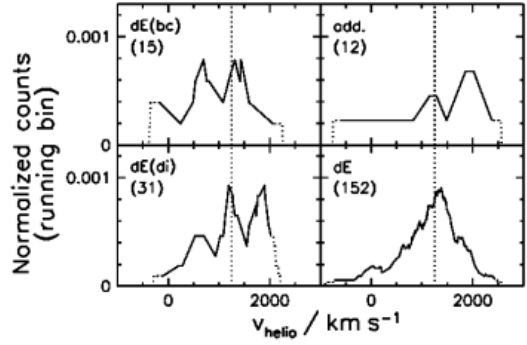}
  \caption{Velocity distribution.
      Distribution of available heliocentric velocities, taken from the
      NASA/IPAC Extragalactic Database (NED). The dE(bc)s of the main
      sample are shown in the top left panel, the galaxies of the
      additional sample in the top right panel, the dE(di)s from Paper I
      in the bottom left panel, and the remaining dEs (excluding dE(bc)s
      and dE(di)s) in the bottom right panel.
      Only galaxies with certain cluster
      membership are included. The vertical dotted
      line marks the value of $v_{\rm helio} = 1248$ km s$^{-1}$, which is
      the median value of all 194 available velocities for early-type
      dwarfs. The data are shown as running histogram with a bin-width
      of $366 $km s$^{-1}$, corresponding to the semi-interquartile range of
      these 194 velocities. Numbers in brackets are the number of
      galaxies included in the respective panel. 
      Beyond the last data point on each side, the histograms
      are plotted with dotted instead of solid lines
  }
  \label{fig:velo}
\end{figure}

Heliocentric velocities are available for 194 early-type dwarfs of our 
full dE sample that are certain Virgo cluster members (see
Sect.~\ref{sec:sub_velodata}).
We present these data in
Fig.~\ref{fig:velo} for the dE(bc)s, the dE(di)s, and the remaining dEs (i.e.,
  dE(bc)s and dE(di)s excluded). Each panel shows a
running histogram with a bin-width of $366$ km s$^{-1}$, which corresponds to
the semi-interquartile range of all 194 velocities. Their median value,
 $v_{\rm helio} = 1248$ km s$^{-1}$, is given as a dotted vertical line in each
panel for comparison.

The dE(bc)s of the main sample show a relatively large deviation from the
overall median value: their median velocity
is only $v_{\rm helio} = 802$ km s$^{-1}$. However, since the semi-interquartile
range is large ($\Delta v_{\rm helio} = 463$ km s$^{-1}$), and since only 15
objects are included in this sample, we cannot state whether this constitutes a significant
difference. Instead, it might be more promising to compare the shapes
of the distributions. The sample of ordinary dEs shows a close-to-symmetric
distribution, which would be expected for a relaxed population.
In contrast, the distribution of dE(bc)s and dE(di)s are asymmetric and less
smooth. Both display side peaks, indicative of an infalling
population \citep{tul84,conI}. This would be consistent with their
spatial distribution, which
shows no central concentration within the cluster,
and with their distribution with projected local density, which follows
that of the irregular cluster galaxies.

%________________________________________________________________

\section{Discussion}
\label{sec:discuss}

The SDSS enables for the first time a systematic study of several hundred dEs
in the Virgo cluster with optical imaging and partly with spectroscopy. These
data provide the basis for this series of papers, which aims at
disentangling the various subpopulations of early-type dwarfs and uncovering
their evolutionary histories. In Paper I, 
we found that dEs with disk features (dE(di)s) constitute a
disk-shaped, unrelaxed dE population that is clearly different from
classical dwarf ellipticals. In the current paper, we focus on another
conspicuous feature that is common to several dEs: a blue center caused by
recent or ongoing central star formation. We have shown that these dE(bc)s
constitute a significant fraction (more than 15\%) of bright dEs in the
Virgo cluster. This number declines to almost zero beyond $\mb > 16$,
which is most likely a real decline and is not mainly due to signal-to-noise
effects.

Two
dE(bc)s of the main sample are detected in HI, with gas-to-baryonic mass
fractions of 1\% and $3-6$\%, respectively. These values and the upper
limits for several other dE(bc)s suggest an average gas content larger
than that of ordinary dEs, but lower than that of dwarf irregulars. Note that
we do not 
know where in the galaxies the gas is located: if it were concentrated in
the central region, it would probably be able to fuel star formation for a
longer time than if the gas were distributed homogeneously.

As soon as star formation
has ceased, each dE(bc)'s color -- which provided the
basis of its selection -- will become indistinguishable from that of an
ordinary dE within $\sim$ a Gigayear or less (see Sect.~\ref{sec:sub_color}).
However, the statistical properties of the dE(bc) \emph{population}
are unlike those of ordinary dEs: the projected
spatial distribution and the flattening distribution of the dE(bc)s are
similar to those of the dE(di)s and are different from those of dEs that have no
blue centers or disk features. Both the dE(bc) and dE(di) population show no central
clustering, which, along with the side-peaks of their velocity
distributions, hints towards fairly recent infall.
How long ago this infall could have taken place depends on relaxation
timescales. \citet{conI} derived a two-body relaxation time for the Virgo dEs of
much more than a Hubble time. Even violent relaxation, which might only
apply for the case of merging groups or subclusters, 
would take at least a few crossing times $t_{\rm cr}$, with $t_{\rm
  cr}\approx 1.7$ Gyr for the Virgo cluster \citep{bos06}. Thus, a dE
population built out of infalling galaxies remains in an
unrelaxed, non-virialized state for many Gyr. We conclude
that the dE(bc)s most likely formed through infall of progenitor
galaxies.

The shape of the dE(bc)s, as deduced from their flattening distribution,
is hardly consistent with their being spheroidal objects, and instead
implies that the dE(bc)s are rather thick disks, i.e.\ oblate-shaped
objects with intrinsic axial ratios around $\sim$0.4. This is only
somewhat thicker than the dE(di)s, for which we derived an axial ratio of
$\sim$0.35 in Paper I. How could these disk-shaped dEs be produced?

\subsection{Formation scenarios}
\label{sec:sub_discuss_formation}

In the galaxy
harassment scenario \citep{moo96}, an infalling late-type disk galaxy
gets transformed into a dE through high-speed encounters with massive
cluster galaxies. This obviously leads to an increase in the axial
ratio during the transformation process, since a disk is
converted into a spheroid. However, a thick stellar disk may survive
and lead to lenticular systems \citep{moo96,mas05}. These may form a
bar and spiral features, and retain them for some time, depending on the
tidal heating of the galaxy \citep{mas05}. Galaxy harassment thus
appears to be a plausible scenario to explain the formation of
disk-shaped dEs, and of dE(di)s in particular. Moreover, it predicts gas to be
funneled to the center and form a density excess there
\citep{moo98}. While Moore et al.\ compare this to the presence of a
nucleus in many dEs, they admit that their simulations ``were not
designed to probe the inner 200 pc''. The radii of many of
the blue central regions of the dE(bc)s are only slightly larger than
this value (see Figs.~\ref{fig:pics1a} to \ref{fig:pics1c});
consequently, the central gas density excess could well 
explain the blue centers. The harassment scenario thus
describes a possible evolution of an infalling late-type disk galaxy to a dE(bc)
and to a dE(di).

Another mechanism to form disk-shaped dEs could be ram-pressure
stripping \citep[e.g.][]{gun72} of dwarf irregulars. Depending on
galaxy mass, the gas might be significantly removed except around the
central region \citep{mor00}, which would seem to be consistent with
the central star formation of the dE(bc)s.
As shown by
\citet{vZe04a}, several dEs have significant rotation, and could thus
be the descendants of dIrrs, which are also known to be mostly
rotationally supported, at least at the luminosities considered
here. However, apart from the problems with this scenario
discussed in Sect.~\ref{sec:intro}, like the metallicity offset between
dEs and dIrrs or the too strong fading of dIrrs, the flattening
distribution of dIrrs shown by \citet{bin95} is not quite like that of our
dE(bc)s. Instead, dIrrs have a (primary) axial ratio $\ge$0.5. On the
other hand, significant mass loss due to stripped gas 
would be expected to affect also the \emph{stellar} configuration of the
galaxies and could thus possibly account for the difference.

As outlined in Sect.~\ref{sec:intro}, several studies claimed that BCDs
might be progenitors of dEs. Their flattening distribution, as analyzed
by \citet{bin95}, is somewhat more like that of our dE(bc)s, though
still slightly rounder. These BCDs behave similarly to the galaxies of our 
additional sample -- which were mostly classified as (candidate) BCDs,
but were selected only if their appearance was similar to the
dE(bc)s. Overall, they have more extended blue regions, stronger emission 
lines, and clearly a higher gas content than the dE(bc)s of the main
sample. They are also dominated by an underlying old
population -- only one out of seven has $M_{\rm old}<90$\% -- and they
have fairly regular outer shapes. Their spatial distribution also hints
at an unrelaxed population, and their velocity distribution is
asymmetric.

Tidally induced star formation in dIrrs \citep{dav88} might be able to
link BCDs and dEs, and could at the same time overcome the problems of the
ram-pressure stripping scenario. The initially lower metallicity and
surface brightness of a dIrr are increased by several bursts of star
formation, during which the galaxy appears as blue compact dwarf
(BCD). After that, it fades to become a dE. The last star
formation burst might occur in the central region of the dwarf
\citep{dav88}, consistent with the appearance of the dE(bc)s. 

\subsection{Presence of nuclei}
\label{sec:sub_discuss_nuclei}

Many late-type spirals -- which act as dE progenitors in the
 harassment scenario -- host a compact nucleus \citep[e.g.][]{boe04}. If
 the nucleus survives the morphological transformation, it would become
 the nucleus of the resulting dE(bc) or dE(di). A second scenario for
 nucleus formation in dEs is late star formation out of
 central gas \citep[e.g.][]{oh00}, which in principle might be taking
 place in the centers of the dE(bc)s.

The fraction of nucleated galaxies among the dE(bc)s is 7/16 (44\%),
which is based on the VCC classification and has been verified by
  us through visual inspection of images and unsharp masks.
We find one more dE(bc) to have ``multiple nuclei'' (VCC 0021), and one
to have a possible nucleus (VCC 1512).
 If those two objects were
counted as being nucleated, the fraction would increase to 56\%. Among
the dE(di)s, 26 out of 36 galaxies are nucleated (72\%). Ordinary dEs
(i.e.\ excluding dE(bc)s and dE(di)s) have a nucleated fraction of also
72\% among the brighter objects ($\mb\le 
16$), which is the magnitude range of most dE(di)s and of all but one
dE(bc). If we assumed a 72\% nucleated fraction for the dE(bc)s as well,
we would expect to find 11.5 nucleated dE(bc)s among our 16 objects,
with a standard deviation $\sigma = 1.8$. Our observed number thus lies
within $1.4 \sigma$ of the expected value if dE(bc)s had the same nucleated
fraction as dEs and dE(di)s, and if the two uncertain objects were
counted as nucleated. Thus, there is no significant difference between
dE(bc)s, dE(di)s, and the other dEs with respect to the presence or
absence of a nucleus. Still, the somewhat smaller number of nuclei in
dE(bc)s could be a hint that nuclei are indeed forming in their centers.
Of the
galaxies in the additional sample, none displays a nucleus, but the
significant amount of central gas and dust might leave the possibility
of a hidden nucleus, or of a nucleus just being formed.

We point out that the ACS Virgo Cluster Survey \citep{acsvir} finds a
much higher frequency of compact stellar nuclei in early-type galaxies
than \citet{vcc} did, primarily due to the much higher
resolution of space-based studies as compared to ground-based ones
\citep{cot05}. However, if we took into account these results, a dE
classified as nucleated in the VCC would then simply be termed \emph{a
  dE with a nucleus bright enough to have been detected by
  \citet{vcc}}.

\subsection{Presence of dE(bc)s in less dense environments}
\label{sec:sub_discuss_ngc205}

Early-type dwarfs with blue centers are not only present in the Virgo
cluster. NGC 205, with its central region of young stars and central
dust clouds \citep{hod73}, is a well
known local example of what we term a dE(bc);
 the same applies to NGC 185 \citep{hod63}.
On the
one hand, NGC 205 might be considered as special case due to 
its clear signs of tidal interaction with M 31 \citep{mat98}. On the
other hand, tidal interaction might not be something 
special but instead something very common and even required for the
formation of dEs, and of dE(bc)s in particular. While galaxy harassment is
negligible in groups \citep{may01}, a similar mechanism is provided
there by
the so-called tidal stirring scenario \citep{may01a}, in which a dIrr
that suffers repeated tidal shocks is transformed into a dE or a dSph.
In this model, the galaxies with higher surface brightness can reach a
central gas density excess like what was described above for
harassment. Tidal stirring might thus provide a consistent explanation
for dE(bc)s in groups.

Contrary to the above interpretation of NGC 185 and NGC 205 as
possible result of tidal stirring, \citet{gu06} recently presented
an apparently isolated dwarf elliptical with a blue center
(IC 225) at a distance of 20.6 Mpc. Its spectrum displays Balmer line
emission, and based on its appearance it would doubtlessly enter our
dE(bc) sample.
The authors base their conclusion about the galaxy's isolation on their
failure to find a potential companion within 30 arcmin, using the
NED. This angular search radius corresponds to 180 kpc at their given
distance. We increased the search radius, and found the small galaxy
group USGC U124 \citep{ram02} to lie at the same distance as IC 225
(within 25 km s$^{-1}$ in radial velocity), and at 
an angular separation of 145 arcmin, or 870 kpc. The brightest group
member (NGC 0936) is an early-type spiral galaxy with
$\Mb\approx-20.5$. Since 870 kpc would seem a rather large distance for
IC 225 to be a bound companion, we test whether a single interaction
could have occured in the past. As an example, if the relative velocity
was 100 km s$^{-1}$, the encounter would have occured 8.5 Gyr ago,
much too long for the current star formation to have been triggered
by it.

While the tidal forces of smaller galaxies closer to IC 225
could possibly affect the galaxy's gas distribution, \citet{bro04}
argued that tidal interactions might not be necessary for activating star
formation, and that instead, the dynamics of gas masses in a dark
matter gravitational potential could be the primary
trigger. Nevertheless, the main problem
would be to explain how an \emph{isolated dE} could have formed at all.
Therefore, if IC 225 is indeed
a truly isolated galaxy, one has to think of other mechanisms
for dE formation than the ones we have discussed.
It also demonstrates along with NGC
185 and NGC 205 that central star formation does not only occur in
cluster dEs, and that, consequently, the mechanisms for dE formation
might be similar in different environments.
In fact, in the Local Group, where we can study galaxies at the highest
angular resolution, many dSphs show star formation histories that
extend over many Gyr. In all of these cases, the younger populations
are more centrally concentrated than the old ones
\citep{har01}. Apart from interactions, this may also be a consequence of
longer-lived gas reservoirs at the centers of these galaxies' potential
wells.

%UZC-SSRS2 Group Catalog (USGC) U124
%galaxy group

Finally, we caution against calling the dE(bc)s ``dE/dIrr
transition types'': this would suggest that every dE(bc) has a dIrr progenitor,
which might not be the case as discussed above. Similarly,
\citet{gre03} argued for Local Group dwarfs that the so-called
dIrr/dSph transition types are plausible progenitors of dSphs, while
dIrrs themselves seem less likely.

%________________________________________________________________

\section{Summary and outlook}
\label{sec:summary}

We have presented a study of Virgo cluster early-type dwarf galaxies
(dEs) with central 
star forming regions, based on photometric and spectroscopic data from
the SDSS DR4. These ``dE(bc)s'' are not rare objects, but they reach a
fraction of more than 15\% among the bright ($\mb < 16$) dEs. Their
spatial distribution
and their distribution with projected local density suggest that
they are an unrelaxed population.
Their flattening distribution is consistent
with them being disk-shaped objects like the dEs with disk substructure
(dE(di)s) identified in 
Paper I. Even in the very 
center, where their colors are bluest, 90\% or more of their mass
belongs to an old stellar population. Thus, they will appear like
ordinary dEs within about one Gigayear after the end of their star
formation. The gas content of the dE(bc)s is lower than in dwarf
irregulars, but probably somewhat higher than in classical dEs,
implying that at least in some dE(bc)s star formation might still
continue for some time.
Plausible formation mechanisms that could explain both the disk shape and
the central star formation of the dE(bc)s are galaxy harassment
\citep{moo96}, which describes the transformation of infalling
late-type disk galaxies into dEs, tidally induced star formation in
dIrrs \citep{dav88}, and possibly also ram-pressure stripping
\citep{gun72} of dIrrs and star formation induced by gas compression
due to ram pressure.

We have started our series of papers on Virgo early-type dwarfs by
describing two sorts of dEs with ``special features'', i.e., dEs with a blue
center (this study) and dEs with disk substructure (Paper I). It is important to
stress that we are not looking at one single population of dEs, with those
features just being some extra ``flavour'', but that these
dE subtypes constitute populations with distinct properties that differ from
the rest of dEs. To complicate the issue even more, ordinary dEs (i.e.,
exluding dE(bc)s and dE(di)s) are probably not a homogeneous population
either, since e.g.\ the clustering properties of nucleated and non-nucleated
dEs differ significantly \citep{bin87}. Whether or not all of these dE
subtypes simply reflect different evolutionary stages of one single class of
galaxy, or whether they are indeed different classes of early-type dwarfs,
will be discussed in detail in a forthcoming paper.

%________________________________________________________________

\acknowledgements
    It is a pleasure to thank B.\ Binggeli for many fruitful
    discussions and a careful reading of the manuscript.
    We thank the referee for useful comments that helped us improve the
    paper.
    We gratefully acknowledge support by the Swiss National Science
    Foundation through grants number 200020-105260, 200020-105535, and
    200021-109616.
    We thank J.\
    Gallagher for stimulating discussions.

    This study would not have been possible without the wealth of publicly
    available data from the SDSS Data Release 4, or without the
    fantastic work of those who created and maintain the SDSS
    webpages.    
    Funding for the SDSS has been provided by the Alfred
    P. Sloan Foundation, the Participating Institutions, the National
    Science Foundation, the U.S. Department of Energy, the National
    Aeronautics and Space Administration, the Japanese Monbukagakusho,
    the Max Planck Society, and the Higher Education Funding Council
    for England. The SDSS Web Site is http://www.sdss.org/. 

    The SDSS is managed by the Astrophysical Research Consortium for
    the Participating Institutions. The Participating Institutions are
    the American Museum of Natural History, Astrophysical Institute
    Potsdam, University of Basel, Cambridge University, Case Western
    Reserve University, University of Chicago, Drexel University,
    Fermilab, the Institute for Advanced Study, the Japan Participation
    Group, Johns Hopkins University, the Joint Institute for Nuclear
    Astrophysics, the Kavli Institute for Particle Astrophysics and
    Cosmology, the Korean Scientist Group, the Chinese Academy of
    Sciences (LAMOST), Los Alamos National Laboratory, the
    Max-Planck-Institute for Astronomy (MPIA), the Max-Planck-Institute
    for Astrophysics (MPA), New Mexico State University, Ohio State
    University, University of Pittsburgh, University of Portsmouth,
    Princeton University, the United States Naval Observatory, and the
    University of Washington. 

    This research has made use of NASA's Astrophysics Data
    System Bibliographic Services, and of the NASA/IPAC Extragalactic
    Database (NED) which is operated by the Jet Propulsion Laboratory,
    California Institute of Technology, under contract with the
    National Aeronautics and Space Administration.
    This research has also made use of the GOLDMine Database.

%________________________________________________________________

%\bibliography{virgodE}
%\bibliographystyle{aa}
%\bibliography{velorefs}
%\bibliographystyle{aa}
%\bibliography{bluecore4}

\begin{thebibliography}

\bibitem[{{Adelman-McCarthy} {et~al.}(2006){Adelman-McCarthy}, {Ag{\"u}eros},
  {Allam}, {Anderson}, {Anderson}, {Annis}, {Bahcall}, {Baldry}, {Barentine},
  {Berlind}, {Bernardi}, {Blanton}, {Boroski}, {Brewington}, {Brinchmann},
  {Brinkmann}, {Brunner}, {Budav{\'a}ri}, {Carey}, {Carr}, {Castander},
  {Connolly}, {Csabai}, {Czarapata}, {Dalcanton}, {Doi}, {Dong}, {Eisenstein},
  {Evans}, {Fan}, {Finkbeiner}, {Friedman}, {Frieman}, {Fukugita}, {Gillespie},
  {Glazebrook}, {Gray}, {Grebel}, {Gunn}, {Gurbani}, {de Haas}, {Hall},
  {Harris}, {Harvanek}, {Hawley}, {Hayes}, {Hendry}, {Hennessy}, {Hindsley},
  {Hirata}, {Hogan}, {Hogg}, {Holmgren}, {Holtzman}, {Ichikawa}, {Ivezi{\'c}},
  {Jester}, {Johnston}, {Jorgensen}, {Juri{\'c}}, {Kent}, {Kleinman}, {Knapp},
  {Kniazev}, {Kron}, {Krzesinski}, {Kuropatkin}, {Lamb}, {Lampeitl}, {Lee},
  {Leger}, {Lin}, {Long}, {Loveday}, {Lupton}, {Margon},
  {Mart{\'{\i}}nez-Delgado}, {Mandelbaum}, {Matsubara}, {McGehee}, {McKay},
  {Meiksin}, {Munn}, {Nakajima}, {Nash}, {Neilsen}, {Newberg}, {Newman},
  {Nichol}, {Nicinski}, {Nieto-Santisteban}, {Nitta}, {O'Mullane}, {Okamura},
  {Owen}, {Padmanabhan}, {Pauls}, {Peoples}, {Pier}, {Pope}, {Pourbaix},
  {Quinn}, {Richards}, {Richmond}, {Rockosi}, {Schlegel}, {Schneider},
  {Schroeder}, {Scranton}, {Seljak}, {Sheldon}, {Shimasaku}, {Smith}, {Smol{\v
  c}i{\'c}}, {Snedden}, {Stoughton}, {Strauss}, {SubbaRao}, {Szalay},
  {Szapudi}, {Szkody}, {Tegmark}, {Thakar}, {Tucker}, {Uomoto}, {Vanden Berk},
  {Vandenberg}, {Vogeley}, {Voges}, {Vogt}, {Walkowicz}, {Weinberg}, {West},
  {White}, {Xu}, {Yanny}, {Yocum}, {York}, {Zehavi}, {Zibetti}, \&
  {Zucker}}]{sdssdr4}
{Adelman-McCarthy}, J.~K., {Ag{\"u}eros}, M.~A., {Allam}, S.~S., {et~al.} 2006,
  \apjs, 162, 38

\bibitem[{{Binggeli}(1985)}]{bin85}
{Binggeli}, B. 1985, in Star-Forming Dwarf Galaxies and Related Objects, ed.
  D.~{Kunth}, T.~X. {Thuan}, \& J.~{Tran Thanh van}, 53

\bibitem[Binggeli \& Cameron(1991)]{bin91} Binggeli, B., \& Cameron L., M. 1991,
      A\&A, 252, 27

\bibitem[{{Binggeli} \& {Cameron}(1993)}]{bin93}
{Binggeli}, B. \& {Cameron}, L.~M. 1993, \aaps, 98, 297

\bibitem[{{Binggeli} \& {Popescu}(1995)}]{bin95}
{Binggeli}, B. \& {Popescu}, C.~C. 1995, \aap, 298, 63

\bibitem[{{Binggeli} {et~al.}(1993){Binggeli}, {Popescu}, \&
  {Tammann}}]{virgokin}
{Binggeli}, B., {Popescu}, C.~C., \& {Tammann}, G.~A. 1993, \aaps, 98, 275

\bibitem[{{Binggeli} {et~al.}(1985){Binggeli}, {Sandage}, \& {Tammann}}]{vcc}
{Binggeli}, B., {Sandage}, A., \& {Tammann}, G.~A. 1985, \aj, 90, 1681

\bibitem[{{Binggeli} {et~al.}(1987){Binggeli}, {Tammann}, \& {Sandage}}]{bin87}
{Binggeli}, B., {Tammann}, G.~A., \& {Sandage}, A. 1987, \aj, 94, 251

\bibitem[{{B{\"o}ker} {et~al.}(2004){B{\"o}ker}, {Walcher}, {Rix},
  {H{\"a}ring}, {Schinnerer}, {Sarzi}, {van der Marel}, {Ho}, {Shields},
  {Lisenfeld}, \& {Laine}}]{boe04}
{B{\"o}ker}, T., {Walcher}, C.~J., {Rix}, H.~W., {et~al.} 2004, in ASP Conf.
  Ser. 322: The Formation and Evolution of Massive Young Star Clusters, ed.
  H.~J.~G.~L.~M. {Lamers}, L.~J. {Smith}, \& A.~{Nota}, 39

\bibitem[{{Boselli} \& {Gavazzi}(2006)}]{bos06}
{Boselli}, A. \& {Gavazzi}, G. 2006, \pasp, 118, 517

\bibitem[{{Boselli} {et~al.}(2002){Boselli}, {Iglesias-P{\'a}ramo},
  {V{\'{\i}}lchez}, \& {Gavazzi}}]{bos02}
{Boselli}, A., {Iglesias-P{\'a}ramo}, J., {V{\'{\i}}lchez}, J.~M., \&
  {Gavazzi}, G. 2002, \aap, 386, 134

\bibitem[{{Bothun} {et~al.}(1986){Bothun}, {Mould}, {Caldwell}, \&
  {MacGillivray}}]{bot86}
{Bothun}, G.~D., {Mould}, J.~R., {Caldwell}, N., \& {MacGillivray}, H.~T. 1986,
  \aj, 92, 1007

\bibitem[{{Brosch} {et~al.}(2004){Brosch}, {Almoznino}, \& {Heller}}]{bro04}
{Brosch}, N., {Almoznino}, E., \& {Heller}, A.~B. 2004, \mnras, 349, 357

  \bibitem[Bruzual \& Charlot(1993)]{bruzual_93} Bruzual A., G., \& Charlot, S. 1993,
      ApJ, 405, 538

  \bibitem[Bruzual \& Charlot(2000)]{bruzual_00} Bruzual A., G., \& Charlot, S. 2000,
      Galaxy isochrone spectral synthesis evolution library (private communication)

  \bibitem[Bruzual \& Charlot(2003)]{bruzual_03} Bruzual A., G., \& Charlot, S. 2003,
      MNRAS, 344, 1000

\bibitem[{{Caldwell} {et~al.}(2003){Caldwell}, {Rose}, \&
  {Concannon}}]{2003AJ....125.2891C}
{Caldwell}, N., {Rose}, J.~A., \& {Concannon}, K.~D. 2003, \aj, 125, 2891

  \bibitem[Calzetti et al.(2000)]{calzetti} Calzetti, D., Armus, L., Bohlin, R. C., et al. 2000,
      ApJ, 533, 682

  \bibitem[Charbonnel et al.(1993)]{charbonnel} Charbonnel, C., Maeder, A., Schaller, G., et al. 1993,
      A\&AS, 101, 415

  \bibitem[Charlot \& Bruzual(1991)]{charlot_91} Charlot, S., \& Bruzual A., G. 1991,
      ApJ, 367, 126

  \bibitem[Cid Fernandes et al.(2003)]{cidfernandes} Cid Fernandes, R., Le\~{a}o, J. R. S., \& Lacerda, R. R. 2003,
      MNRAS, 340, 29

\bibitem[{{Conselice} {et~al.}(2001){Conselice}, {Gallagher}, \&
  {Wyse}}]{conI}
{Conselice}, C.~J., {Gallagher}, III, J.~S., \& {Wyse}, R.~F.~G. 2001, \apj,
  559, 791

\bibitem[{{Conselice} {et~al.}(2003){Conselice}, {O'Neil}, {Gallagher}, \&
  {Wyse}}]{conIV}
{Conselice}, C.~J., {O'Neil}, K., {Gallagher}, J.~S., \& {Wyse}, R.~F.~G. 2003,
  \apj, 591, 167

\bibitem[{{C{\^o}t{\'e}}(2005)}]{cot05}
{C{\^o}t{\'e}}, P. 2005, in IAU Colloq. 198: Near-field cosmology with dwarf
  elliptical galaxies, ed. H.~{Jerjen} \& B.~{Binggeli} (Cambridge: CUP), 269--276

\bibitem[{{C{\^o}t{\'e}} {et~al.}(2004){C{\^o}t{\'e}}, {Blakeslee},
  {Ferrarese}, {Jord{\'a}n}, {Mei}, {Merritt}, {Milosavljevi{\'c}}, {Peng},
  {Tonry}, \& {West}}]{acsvir}
{C{\^o}t{\'e}}, P., {Blakeslee}, J.~P., {Ferrarese}, L., {et~al.} 2004, \apjs,
  153, 223

  \bibitem[Cuisinier et al.(2006)]{cuisinier_06} Cuisinier, F.,
      Westera, P., Telles, E., \& Buser, R. 2006, A\&A, 423, 133

\bibitem[{{Davies} \& {Phillipps}(1988)}]{dav88}
{Davies}, J.~I. \& {Phillipps}, S. 1988, \mnras, 233, 553

\bibitem[{{de Vaucouleurs} {et~al.}(1991){de Vaucouleurs}, {de Vaucouleurs},
  {Corwin}, {Buta}, {Paturel}, \& {Fouque}}]{1991trcb.book.....D}
{de Vaucouleurs}, G., {de Vaucouleurs}, A., {Corwin}, Jr., H.~G., {et~al.}
  1991, {Third Reference Catalogue of Bright Galaxies} (Volume 1-3, XII, 2069
  pp., Springer-Verlag Berlin Heidelberg New York)

\bibitem[{{Dressler}(1980)}]{dre80}
{Dressler}, A. 1980, \apj, 236, 351

\bibitem[{{Drinkwater} {et~al.}(1996){Drinkwater}, {Currie}, {Young}, {Hardy},
  \& {Yearsley}}]{dri96}
{Drinkwater}, M.~J., {Currie}, M.~J., {Young}, C.~K., {Hardy}, E., \&
  {Yearsley}, J.~M. 1996, \mnras, 279, 595

\bibitem[Fagotto et al.(1994)]{fagotto2}
  Fagotto F., Bressan A., Bertelli G., \& Chiosi C., 1994 A\&AS 105, 29

\bibitem[{{Falco} {et~al.}(1999){Falco}, {Kurtz}, {Geller}, {Huchra}, {Peters},
  {Berlind}, {Mink}, {Tokarz}, \& {Elwell}}]{1999PASP..111..438F}
{Falco}, E.~E., {Kurtz}, M.~J., {Geller}, M.~J., {et~al.} 1999, \pasp, 111, 438

\bibitem[{{Ferguson} \& {Binggeli}(1994)}]{fer94}
{Ferguson}, H.~C. \& {Binggeli}, B. 1994, \aapr, 6, 67


\bibitem[{{Ferrarese} {et~al.}(2006){Ferrarese}, {Cote}, {Jordan}, {Peng},
  {Blakeslee}, {Piatek}, {Mei}, {Merritt}, {Milosavljevic}, {Tonry}, \&
  {West}}]{fer06}
{Ferrarese}, L., {Cote}, P., {Jordan}, A., {et~al.} 2006, \apjs, 164, 334

\bibitem[{{Ferrarese} {et~al.}(2000){Ferrarese}, {Mould}, {Kennicutt},
  {Huchra}, {Ford}, {Freedman}, {Stetson}, {Madore}, {Sakai}, {Gibson},
  {Graham}, {Hughes}, {Illingworth}, {Kelson}, {Macri}, {Sebo}, \&
  {Silbermann}}]{fer00}
{Ferrarese}, L., {Mould}, J.~R., {Kennicutt}, Jr., R.~C., {et~al.} 2000, \apj,
  529, 745

  \bibitem[Fluks et al.(1994)]{fluks} Fluks, M. A., Plez, B., Th{\'e}, P. S., et al. 1994,
      A\&AS, 105, 311

\bibitem[{{Gavazzi} {et~al.}(2000){Gavazzi}, {Bonfanti}, {Pedotti}, {Boselli},
  \& {Carrasco}}]{2000A&AS..146..259G}
{Gavazzi}, G., {Bonfanti}, C., {Pedotti}, P., {Boselli}, A., \& {Carrasco}, L.
  2000, \aaps, 146, 259

\bibitem[{{Gavazzi} {et~al.}(2003){Gavazzi}, {Boselli}, {Donati}, {Franzetti},
  \& {Scodeggio}}]{goldmine}
{Gavazzi}, G., {Boselli}, A., {Donati}, A., {Franzetti}, P., \& {Scodeggio}, M.
  2003, \aap, 400, 451

\bibitem[{{Gavazzi} {et~al.}(2005){Gavazzi}, {Boselli}, {van Driel}, \&
  {O'Neil}}]{gav05}
{Gavazzi}, G., {Boselli}, A., {van Driel}, W., \& {O'Neil}, K. 2005, \aap, 429,
  439

\bibitem[{{Gavazzi} {et~al.}(2004){Gavazzi}, {Zaccardo}, {Sanvito}, {Boselli},
  \& {Bonfanti}}]{2004A&A...417..499G}
{Gavazzi}, G., {Zaccardo}, A., {Sanvito}, G., {Boselli}, A., \& {Bonfanti}, C.
  2004, \aap, 417, 499

\bibitem[{{Geha} {et~al.}(2003){Geha}, {Guhathakurta}, \& {van der
  Marel}}]{geh03}
{Geha}, M., {Guhathakurta}, P., \& {van der Marel}, R.~P. 2003, \aj, 126, 1794

  \bibitem[Girardi et al.(1996)]{girardi} Girardi, L., Bressan, A., Chiosi, C., Bertelli, G., \& Nasi, E. 1996,
      A\&AS, 117, 113

  \bibitem[Girardi et al.(2000)]{girardi_00} Girardi, L., Bressan, A., Bertelli, G., \& Chiosi, C. 2000,
      A\&AS, 141, 371 (``Padova 2000'' isochrones)

\bibitem[{{Grebel}(2001)}]{gre01}
{Grebel}, E.~K. 2001, Astrophysics and Space Science Supplement, 277, 231

\bibitem[{{Grebel} {et~al.}(2003){Grebel}, {Gallagher}, \& {Harbeck}}]{gre03}
{Grebel}, E.~K., {Gallagher}, J.~S., \& {Harbeck}, D. 2003, \aj, 125, 1926

\bibitem[{{Grogin} {et~al.}(1998){Grogin}, {Geller}, \&
  {Huchra}}]{1998ApJS..119..277G}
{Grogin}, N.~A., {Geller}, M.~J., \& {Huchra}, J.~P. 1998, \apjs, 119, 277

  \bibitem[Gu et al.(2006)]{gu06} Gu, Q., Zhao, Y., Shi, L., Peng, Z., \& Luo, X. 2006,
      AJ, 131, 806

\bibitem[{{Gunn} \& {Gott}(1972)}]{gun72}
{Gunn}, J.~E. \& {Gott}, J.~R.~I. 1972, \apj, 176, 1

\bibitem[{{Gunn} {et~al.}(1998){Gunn}, {Carr}, {Rockosi}, {Sekiguchi}, {Berry},
  {Elms}, {de Haas}, {Ivezi{\'c}}, {Knapp}, {Lupton}, {Pauls}, {Simcoe},
  {Hirsch}, {Sanford}, {Wang}, {York}, {Harris}, {Annis}, {Bartozek},
  {Boroski}, {Bakken}, {Haldeman}, {Kent}, {Holm}, {Holmgren}, {Petravick},
  {Prosapio}, {Rechenmacher}, {Doi}, {Fukugita}, {Shimasaku}, {Okada}, {Hull},
  {Siegmund}, {Mannery}, {Blouke}, {Heidtman}, {Schneider}, {Lucinio}, \&
  {Brinkman}}]{sdsscamera}
{Gunn}, J.~E., {Carr}, M., {Rockosi}, C., {et~al.} 1998, \aj, 116, 3040

\bibitem[{{Guzman} {et~al.}(1996){Guzman}, {Koo}, {Faber}, {Illingworth},
  {Takamiya}, {Kron}, \& {Bershady}}]{guz96}
{Guzman}, R., {Koo}, D.~C., {Faber}, S.~M., {et~al.} 1996, \apjl, 460, L5

\bibitem[{{Harbeck} {et~al.}(2001){Harbeck}, {Grebel}, {Holtzman},
  {Guhathakurta}, {Brandner}, {Geisler}, {Sarajedini}, {Dolphin},
  {Hurley-Keller}, \& {Mateo}}]{har01}
{Harbeck}, D., {Grebel}, E.~K., {Holtzman}, J., {et~al.} 2001, \aj, 122, 3092

\bibitem[{{Hodge}(1973)}]{hod73}
{Hodge}, P.~W. 1973, \apj, 182, 671

\bibitem[{{Hodge}(1963)}]{hod63}
{Hodge}, P.~W. 1963, \aj, 68, 691

\bibitem[{{Huchtmeier} \& {Richter}(1989)}]{huc89}
{Huchtmeier}, W.~K. \& {Richter}, O.-G. 1989, \aap, 210, 1

\bibitem[{{Huchtmeier} \& {Richter}(1986)}]{huc86}
{Huchtmeier}, W.~K. \& {Richter}, O.-G. 1986, \aaps, 64, 111

\bibitem[{{Jerjen} {et~al.}(2000){Jerjen}, {Kalnajs}, \& {Binggeli}}]{jer00a}
{Jerjen}, H., {Kalnajs}, A., \& {Binggeli}, B. 2000, \aap, 358, 845

\bibitem[{{Johnson} {et~al.}(1997){Johnson}, {Lawrence}, {Terlevich}, \&
  {Carter}}]{joh97}
{Johnson}, R.~A., {Lawrence}, A., {Terlevich}, R., \& {Carter}, D. 1997,
  \mnras, 287, 333

\bibitem[{{Knezek} {et~al.}(1999){Knezek}, {Sembach}, \& {Gallagher}}]{kne99}
{Knezek}, P.~M., {Sembach}, K.~R., \& {Gallagher}, J.~S. 1999, \apj, 514, 119

\bibitem[{{Kniazev} {et~al.}(2004){Kniazev}, {Pustilnik}, {Grebel}, {Lee}, \&
  {Pramskij}}]{kni04}
{Kniazev}, A.~Y., {Pustilnik}, S.~A., {Grebel}, E.~K., {Lee}, H., \&
  {Pramskij}, A.~G. 2004, \apjs, 153, 429

  \bibitem[Kong et al.(2003)]{kong} Kong, X., Charlot, S., Weiss, A., \& Cheng, F. 2003,
      A\&A, 403, 877

  \bibitem[Le Borgne et al.(2003)]{leborgne} Le Borgne, J.-F., Bruzual A., G., Pell{\'o}, R., et al. 2003,
      A\&A, 402, 433

  \bibitem[Leitherer et al.(1999)]{leitherer} Leitherer, C.,  Schaerer, D., Goldader, J. D., et al. 1999,
      ApJS, 123, 3

  \bibitem[Lejeune et al.(1997)]{lejeune_97} Lejeune, T.,  Cuisinier, F., \& Buser, R. 1997,
      A\&AS, 125, 229

  \bibitem[Lejeune et al.(1998)]{lejeune_98} Lejeune, T., Cuisinier, F., \& Buser, R. 1998,
      A\&AS, 130, 65

\bibitem[{{Lilly} \& {Fritze-v.~Alvensleben}(2003)}]{lil03}
{Lilly}, T. \& {Fritze-v.~Alvensleben}, U. 2004, in IAU Symp. 221:
Star Formation at High Angular Resolution, ed. M.~{Burton},
R.~{Jayawardhana}, \& T.~{Bourke}, 80

\bibitem[{{Lisker} {et~al.}(2005){Lisker}, {Grebel}, \& {Binggeli}}]{lis05}
{Lisker}, T., {Grebel}, E.~K., \& {Binggeli}, B. 2005, in IAU Colloq. 198:
  Near-field cosmology with dwarf elliptical galaxies, ed. H.~{Jerjen} \&
  B.~{Binggeli} (Cambridge: CUP), 311

\bibitem[{{Lisker} {et~al.}(2006){Lisker}, {Grebel}, \& {Binggeli}}]{p1}
{Lisker}, T., {Grebel}, E.~K., \& {Binggeli}, B. 2006, \aj, 132, 497,
  {\it Paper I}

\bibitem[{{Maraston}(2005)}]{mar05}
{Maraston}, C. 2005, \mnras, 362, 799

\bibitem[{{Mastropietro} {et~al.}(2005){Mastropietro}, {Moore}, {Mayer},
  {Debattista}, {Piffaretti}, \& {Stadel}}]{mas05}
{Mastropietro}, C., {Moore}, B., {Mayer}, L., {et~al.} 2005, \mnras, 364, 607

\bibitem[{{Mateo}(1998)}]{mat98}
{Mateo}, M.~L. 1998, \araa, 36, 435

\bibitem[{{Mayer} {et~al.}(2001{\natexlab{b}}){Mayer}, {Governato}, {Colpi}, {Moore},
  {Quinn}, {Wadsley}, {Stadel}, \& {Lake}}]{may01}
{Mayer}, L., {Governato}, F., {Colpi}, M., {et~al.} 2001{\natexlab{b}}, \apj, 559, 754

\bibitem[{{Mayer} {et~al.}(2001{\natexlab{a}}){Mayer}, {Governato}, {Colpi}, {Moore},
  {Quinn}, {Wadsley}, {Stadel}, \& {Lake}}]{may01a}
{Mayer}, L., {Governato}, F., {Colpi}, M., {et~al.} 2001{\natexlab{a}}, \apjl, 547, L123

  \bibitem[Meynet et al.(1994)]{meynet} Meynet, G., Maeder, A., Schaller, G., Schaerer, D., \& Charbonnel, C. 1994,
      A\&AS, 103, 97

\bibitem[{{Moore} {et~al.}(1996){Moore}, {Katz}, {Lake}, {Dressler}, \&
  {Oemler}}]{moo96}
{Moore}, B., {Katz}, N., {Lake}, G., {Dressler}, A., \& {Oemler}, A. 1996,
  \nat, 379, 613

\bibitem[{{Moore} {et~al.}(1998){Moore}, {Lake}, \& {Katz}}]{moo98}
{Moore}, B., {Lake}, G., \& {Katz}, N. 1998, \apj, 495, 139

\bibitem[{{Mori} \& {Burkert}(2000)}]{mor00}
{Mori}, M. \& {Burkert}, A. 2000, \apj, 538, 559

  \bibitem[Oh \& Lin(2000)]{oh00} Oh, K.S. \& Lin, D., N., C. 2000,
    ApJ, 543, 620

  \bibitem[Osterbrock(1989)]{osterbrock} Osterbrock, D. E. 1989,
      Astrophysics of Gaseous Nebulae and Active Galactic Nuclei
      (Mill Valley: University Science Books)

\bibitem[{{Papaderos} {et~al.}(1996){Papaderos}, {Loose}, {Fricke}, \&
  {Thuan}}]{pap96}
{Papaderos}, P., {Loose}, H.-H., {Fricke}, K.~J., \& {Thuan}, T.~X. 1996, \aap,
  314, 59
%
%  \bibitem[Pickles(1998)]{pickles} Pickles, A. J. 1998,
%      PASP, 110 863

\bibitem[{{Pier} {et~al.}(2003){Pier}, {Munn}, {Hindsley}, {Hennessy}, {Kent},
  {Lupton}, \& {Ivezi{\' c}}}]{sdssastrometry}
{Pier}, J.~R., {Munn}, J.~A., {Hindsley}, R.~B., {et~al.} 2003, \aj, 125, 1559

\bibitem[{{Rakos} \& {Schombert}(2004)}]{rak04}
{Rakos}, K. \& {Schombert}, J. 2004, \aj, 127, 1502

\bibitem[{{Ramella} {et~al.}(2002){Ramella}, {Geller}, {Pisani}, \& {da
  Costa}}]{ram02}
{Ramella}, M., {Geller}, M.~J., {Pisani}, A., \& {da Costa}, L.~N. 2002, \aj,
  123, 2976

\bibitem[{{Richer} {et~al.}(1998){Richer}, {McCall}, \& {Stasinska}}]{ric98}
{Richer}, M., {McCall}, M.~L., \& {Stasinska}, G. 1998, \aap, 340, 67

\bibitem[{{Ryden} \& {Terndrup}(1994)}]{ryd94}
{Ryden}, B.~S. \& {Terndrup}, D.~M. 1994, \apj, 425, 43

\bibitem[{{Sandage} \& {Binggeli}(1984)}]{san84}
{Sandage}, A. \& {Binggeli}, B. 1984, \aj, 89, 919

  \bibitem[Schaerer et al.(1993a)]{schaerera} Schaerer, D., Charbonnel, C., Meynet, G., Maeder, A., \& Schaller, G. 1993a,
      A\&AS, 102, 339

  \bibitem[Schaerer et al.(1993b)]{schaererb} Schaerer, D., Meynet, G., Maeder, A., \& Schaller, G. 1993b,
      A\&AS, 98, 523

  \bibitem[Schaller et al.(1992)]{schaller} Schaller, G., Schaerer, D., Meynet, G., \& Maeder, A. 1992,
      A\&AS, 96, 269

\bibitem[{{Schlegel} {et~al.}(1998){Schlegel}, {Finkbeiner}, \&
  {Davis}}]{sch98}
{Schlegel}, D.~J., {Finkbeiner}, D.~P., \& {Davis}, M. 1998, \apj, 500, 525

  \bibitem[Schmutz et al.(1992)]{schmutz} Schmutz, W., Leitherer, C., \& Gruenwald, R. 1992,
      PASP, 104, 1164

\bibitem[{{Simien} \& {Prugniel}(2002)}]{2002A&A...384..371S}
{Simien}, F. \& {Prugniel}, P. 2002, \aap, 384, 371

\bibitem[{{Smith} {et~al.}(2002){Smith}, {Tucker}, {Kent}, {Richmond},
  {Fukugita}, {Ichikawa}, {Ichikawa}, {Jorgensen}, {Uomoto}, {Gunn}, {Hamabe},
  {Watanabe}, {Tolea}, {Henden}, {Annis}, {Pier}, {McKay}, {Brinkmann}, {Chen},
  {Holtzman}, {Shimasaku}, \& {York}}]{smi02}
{Smith}, J.~A., {Tucker}, D.~L., {Kent}, S., {et~al.} 2002, \aj, 123, 2121

\bibitem[{{Stoughton} {et~al.}(2002){Stoughton}, {Lupton}, {Bernardi},
  {Blanton}, {Burles}, {Castander}, {Connolly}, {Eisenstein}, {Frieman},
  {Hennessy}, {Hindsley}, {Ivezi{\'c}}, {Kent}, {Knapp}, {Kunszt}, {Lee},
  {Meiksin}, {Munn}, {Newberg}, {Nichol}, {Nicinski}, {Pier}, {Richards},
  {Richmond}, {Schlegel}, {Smith}, {Strauss}, {SubbaRao}, {Szalay}, {Thakar},
  {Tucker}, {Vanden Berk}, {Yanny}, {Adelman}, {Anderson}, {Anderson}, {Annis},
  {Bahcall}, {Bakken}, {Bartelmann}, {Bastian}, {Bauer}, {Berman},
  {B{\"o}hringer}, {Boroski}, {Bracker}, {Briegel}, {Briggs}, {Brinkmann},
  {Brunner}, {Carey}, {Carr}, {Chen}, {Christian}, {Colestock}, {Crocker},
  {Csabai}, {Czarapata}, {Dalcanton}, {Davidsen}, {Davis}, {Dehnen},
  {Dodelson}, {Doi}, {Dombeck}, {Donahue}, {Ellman}, {Elms}, {Evans}, {Eyer},
  {Fan}, {Federwitz}, {Friedman}, {Fukugita}, {Gal}, {Gillespie}, {Glazebrook},
  {Gray}, {Grebel}, {Greenawalt}, {Greene}, {Gunn}, {de Haas}, {Haiman},
  {Haldeman}, {Hall}, {Hamabe}, {Hansen}, {Harris}, {Harris}, {Harvanek},
  {Hawley}, {Hayes}, {Heckman}, {Helmi}, {Henden}, {Hogan}, {Hogg}, {Holmgren},
  {Holtzman}, {Huang}, {Hull}, {Ichikawa}, {Ichikawa}, {Johnston}, {Kauffmann},
  {Kim}, {Kimball}, {Kinney}, {Klaene}, {Kleinman}, {Klypin}, {Korienek},
  {Krolik}, {Kron}, {Krzesi{\'n}ski}, {Lamb}, {Leger}, {Limmongkol},
  {Lindenmeyer}, {Long}, {Loomis}, {Loveday}, {MacKinnon}, {Mannery},
  {Mantsch}, {Margon}, {McGehee}, {McKay}, {McLean}, {Menou}, {Merelli}, {Mo},
  {Monet}, {Nakamura}, {Narayanan}, {Nash}, {Neilsen}, {Newman}, {Nitta},
  {Odenkirchen}, {Okada}, {Okamura}, {Ostriker}, {Owen}, {Pauls}, {Peoples},
  {Peterson}, {Petravick}, {Pope}, {Pordes}, {Postman}, {Prosapio}, {Quinn},
  {Rechenmacher}, {Rivetta}, {Rix}, {Rockosi}, {Rosner}, {Ruthmansdorfer},
  {Sandford}, {Schneider}, {Scranton}, {Sekiguchi}, {Sergey}, {Sheth},
  {Shimasaku}, {Smee}, {Snedden}, {Stebbins}, {Stubbs}, {Szapudi}, {Szkody},
  {Szokoly}, {Tabachnik}, {Tsvetanov}, {Uomoto}, {Vogeley}, {Voges}, {Waddell},
  {Walterbos}, {Wang}, {Watanabe}, {Weinberg}, {White}, {White}, {Wilhite},
  {Wolfe}, {Yasuda}, {York}, {Zehavi}, \& {Zheng}}]{sdssedr}
{Stoughton}, C., {Lupton}, R.~H., {Bernardi}, M., {et~al.} 2002, \aj, 123, 485

\bibitem[{{Strauss} {et~al.}(1992){Strauss}, {Huchra}, {Davis}, {Yahil},
  {Fisher}, \& {Tonry}}]{1992ApJS...83...29S}
{Strauss}, M.~A., {Huchra}, J.~P., {Davis}, M., {et~al.} 1992, \apjs, 83, 29

\bibitem[{{Thuan}(1985)}]{thu85}
{Thuan}, T.~X. 1985, \apj, 299, 881

\bibitem[{{Tody}(1993)}]{iraf}
{Tody}, D. 1993, in ASP Conf. Ser. 52, Astronomical Data Analysis Software and
  Systems II, ed. R. J. Hanisch, R. J. V. Brissenden, and J. Barnes
  (San Francisco: ASP), 173

  \bibitem[Torres-Peimbert et al.(1989)]{torrespeimbert} Torres-Peimbert, S., Peimbert, M., \& Fiero, J. 1989,
      ApJ, 345, 186

\bibitem[{{Tully} \& {Shaya}(1984)}]{tul84}
{Tully}, R.~B. \& {Shaya}, E.~J. 1984, \apj, 281, 31

\bibitem[{{van Driel} {et~al.}(2000){van Driel}, {Ragaigne}, {Boselli},
  {Donas}, \& {Gavazzi}}]{2000A&AS..144..463V}
{van Driel}, W., {Ragaigne}, D., {Boselli}, A., {Donas}, J., \& {Gavazzi}, G.
  2000, \aaps, 144, 463

\bibitem[{{van Zee} {et~al.}(2004){van Zee}, {Skillman}, \& {Haynes}}]{vZe04a}
{van Zee}, L., {Skillman}, E.~D., \& {Haynes}, M.~P. 2004, \aj, 128, 121

\bibitem[{{Vigroux} {et~al.}(1984){Vigroux}, {Souviron}, \& {Vader}}]{vig84}
{Vigroux}, L., {Souviron}, J., \& {Vader}, J.~P. 1984, \aap, 139, L9

  \bibitem[Westera(2001)]{diss} Westera, P. 2001,
      The BaSeL 3.1 models: Metallicity calibration of a theoretical stellar
      spectral library and its application to chemo-dynamical galaxy models,
      PhD thesis, Univ. of Basel, 378 pp.

  \bibitem[Westera et al.(2002)]{paperiii} Westera, P., Lejeune, T., Buser, R., Cuisinier, F., \& Bruzual A., G. 2002,
      A\&A, 381, 524

  \bibitem[Westera et al.(2004)]{westera_04} Westera, P., Cuisinier, F., Telles, E., \& Kehrig, C. 2004,
      A\&A, 423, 133

\bibitem[{{York} {et~al.}(2000){York}, {Adelman}, {Anderson}, {Anderson},
  {Annis}, {Bahcall}, {Bakken}, {Barkhouser}, {Bastian}, {Berman}, {Boroski},
  {Bracker}, {Briegel}, {Briggs}, {Brinkmann}, {Brunner}, {Burles}, {Carey},
  {Carr}, {Castander}, {Chen}, {Colestock}, {Connolly}, {Crocker}, {Csabai},
  {Czarapata}, {Davis}, {Doi}, {Dombeck}, {Eisenstein}, {Ellman}, {Elms},
  {Evans}, {Fan}, {Federwitz}, {Fiscelli}, {Friedman}, {Frieman}, {Fukugita},
  {Gillespie}, {Gunn}, {Gurbani}, {de Haas}, {Haldeman}, {Harris}, {Hayes},
  {Heckman}, {Hennessy}, {Hindsley}, {Holm}, {Holmgren}, {Huang}, {Hull},
  {Husby}, {Ichikawa}, {Ichikawa}, {Ivezi{\'c}}, {Kent}, {Kim}, {Kinney},
  {Klaene}, {Kleinman}, {Kleinman}, {Knapp}, {Korienek}, {Kron}, {Kunszt},
  {Lamb}, {Lee}, {Leger}, {Limmongkol}, {Lindenmeyer}, {Long}, {Loomis},
  {Loveday}, {Lucinio}, {Lupton}, {MacKinnon}, {Mannery}, {Mantsch}, {Margon},
  {McGehee}, {McKay}, {Meiksin}, {Merelli}, {Monet}, {Munn}, {Narayanan},
  {Nash}, {Neilsen}, {Neswold}, {Newberg}, {Nichol}, {Nicinski}, {Nonino},
  {Okada}, {Okamura}, {Ostriker}, {Owen}, {Pauls}, {Peoples}, {Peterson},
  {Petravick}, {Pier}, {Pope}, {Pordes}, {Prosapio}, {Rechenmacher}, {Quinn},
  {Richards}, {Richmond}, {Rivetta}, {Rockosi}, {Ruthmansdorfer}, {Sandford},
  {Schlegel}, {Schneider}, {Sekiguchi}, {Sergey}, {Shimasaku}, {Siegmund},
  {Smee}, {Smith}, {Snedden}, {Stone}, {Stoughton}, {Strauss}, {Stubbs},
  {SubbaRao}, {Szalay}, {Szapudi}, {Szokoly}, {Thakar}, {Tremonti}, {Tucker},
  {Uomoto}, {Vanden Berk}, {Vogeley}, {Waddell}, {Wang}, {Watanabe},
  {Weinberg}, {Yanny}, \& {Yasuda}}]{sdsstechn}
{York}, D.~G., {Adelman}, J., {Anderson}, J.~E., {et~al.} 2000, \aj, 120, 1579

\bibitem[{{Young} \& {Currie}(1995)}]{1995MNRAS.273.1141Y}
{Young}, C.~K. \& {Currie}, M.~J. 1995, \mnras, 273, 1141

\end{thebibliography}
%\bibliographystyle{aa}

\clearpage

%________________________________________________________________

\begin{deluxetable}{lllllll}
%\tabletypesize{\scriptsize}
  \tablecaption{Early-type dwarfs with blue centers. \label{tab:bludE}}
  \tablehead{
    \colhead{VCC}
    & \colhead{$\alpha_{\rm J2000}$} & \colhead{$\delta_{\rm J2000}$}
    & \colhead{Mem.} & \colhead{$m_{\rm B}$} & \colhead{Class}
    & \colhead{Note}\\
    &&&&(mag)&&
  }
  \startdata
\sidehead{Main sample}
0021 & 12$^{\rm h}$10$^{\rm m}$23\fs2 & +10\arcdeg11\arcmin19\arcsec & M & 14.75 & dS0(4) &  \\
0046 & 12 12 11.0 & +12 53 37 & P & 17.00 & dE3?               & S \\
0170 & 12 15 56.3 & +14 26 00 & M & 14.86 & dS0\,pec:            &  \\
0173 & 12 16 00.4 & +08 12 08 & M & 15.00 & dS0(1)?            &  \\
0218 & 12 17 05.4 & +12 17 22 & M & 14.88 & dS0(8),N:          & D2   \\
0278 & 12 18 14.4 & +06 36 14 & P & 15.10 & dS0,Npec           & D1s    \\
0281 & 12 18 15.2 & +13 44 58 & M & 15.30 & dS0 or BCD         & S, H$\alpha\beta$ \\
0308 & 12 18 50.9 & +07 51 43 & M & 14.30 & d:S0$_1$(0),N:     & D1s   \\
0636 & 12 23 21.3 & +15 52 06 & P & 16.44 & dE0,N or S0$_1$(0) &  \\
0674 & 12 23 52.6 & +13 52 58 & M & 18.00 & dE0,N              &  \\
0781 & 12 25 15.2 & +12 42 53 & M & 14.46 & dS0$_3$(5),N:      & S, H$\alpha$ \\
0870 & 12 26 05.4 & +11 48 43 & M & 14.68 & dS0(5),N           & S, H$\alpha\beta$ \\
0951 & 12 26 54.4 & +11 39 50 & M & 14.23 & dE2\,pec,N or dS0(2),N & S, H$\alpha$ \\
0953 & 12 26 54.8 & +13 33 58 & P & 15.70 & dE5?,Npec?         & S \\
1078 & 12 28 11.4 & +09 45 38 & P & 15.30 & dE5\,pec?            & S, H$\alpha$ \\
1488 & 12 33 13.5 & +09 23 51 & M & 14.76 & dE:                & S; {\it H$\alpha$\,lit.} \\
1501 & 12 33 24.7 & +08 41 27 & M & 15.10 & dS0?               &  \\
1512 & 12 33 34.6 & +11 15 43 & M & 15.73 & dS0\,pec             &  \\
1617 & 12 35 30.9 & +06 20 01 & P & 15.00 & d:S0(4)\,pec?        &  \\
1684 & 12 36 39.4 & +11 06 07 & M & 14.87 & dS0(8):            & S, H$\alpha$; D3    \\
1715 & 12 37 28.5 & +08 47 40 & P & 16.20 & dE0\,pec?            &  \\
1779 & 12 39 04.7 & +14 43 52 & M & 14.83 & dS0(6):            &  D3    \\
1912 & 12 42 09.1 & +12 35 48 & M & 14.16 & dS0(8),N           &  \\
\sidehead{Additional sample}
0024 & 12 10 35.7 & +11 45 39 & M & 14.95 & BCD                & S, H$\alpha\beta$ \\
0135 & 12 15 06.9 & +12 00 59 & M & 14.81 & S\,pec / BCD       & S, H$\alpha\beta$ \\
0334 & 12 19 14.2 & +13 52 57 & M & 16.20 & BCD                & \\
0340 & 12 19 22.1 & +05 54 38 & P & 14.43 & BCD or merger      & S, H$\alpha\beta$ \\
0446 & 12 20 57.9 & +06 20 21 & M & 15.50 & Im / BCD:          & \\
0841 & 12 25 47.6 & +14 57 07 & M & 16.70 & BCD                & \\
0890 & 12 26 21.6 & +06 40 11 & P & 16.00 & BCD?               & \\
1175 & 12 29 18.5 & +10 08 13 & M & 15.10 & E5 / S0$_1$(5)     & S, H$\alpha\beta$; M32  \\
1273 & 12 30 17.0 & +09 05 07 & M & 15.25 & ImIII:             & S, H$\alpha\beta$ \\
1437 & 12 32 33.5 & +09 10 25 & M & 15.70 & BCD                & S, H$\alpha\beta$ \\
1499 & 12 33 20.2 & +12 51 04 & M & 14.58 & E3\,pec or S0      & S \\
1955 & 12 43 07.6 & +12 03 00 & M & 14.12 & S\,pec / BCD       & {\it H$\alpha$\,lit.}\\
2007 & 12 44 47.5 & +08 06 25 & M & 15.20 & ImIII / BCD:       & {\it H$\alpha$\,lit.}\\
2033 & 12 46 04.4 & +08 28 34 & M & 14.65 & BCD                & {\it H$\alpha$\,lit.}\\
  \enddata
  \tablecomments{
Cluster membership (column ``Mem.'') is provided by
 \citet{vcc,virgokin}: M\,=\,certain cluster member, P\,=\,possible 
    member. Classification as in the VCC, except for VCC 1488 which has
  been reclassified as probable dE by \citet{geh03} (former class
  E6:). VCC 1175 belongs to the M32-like compact ellipticals
  \citep{vcc}. Notes in the last column are as follows: D\,=\,disk
  identified in Paper I, D1s\,=\,certain disk with spiral arms,
  D2\,=\,probable disk, D3\,=\,possible disk. S\,=\,useful spectrum
  available. H$\alpha\beta$\,=\,spectrum displays Balmer line emission;
  H$\alpha$\,=\,spectrum displays only H$\alpha$ emission. {\it
  H$\alpha$\,lit.}\,=\,H$\alpha$ detection reported by
  \citet{bos02}. Only given if H$\alpha$ is not detected in the SDSS
  spectrum, or if no spectrum was available.
      Units of
    right ascension are hours, minutes, and seconds, and units of
    declination are degrees, arcminutes, and arcseconds.
    }
\end{deluxetable}

%
%\begin{deluxetable}{lll}
%   \tablecaption{Evolutionary tracks and stellar libraries used in the
%     SSP libraries.\label{tab:libraries}}
%   \tablehead{SSP library   & tracks & stellar library}
%   \startdata
%   ``BC99''      & Padova 2000 & BaSeL 3.1 ``Padova 2000'' \\
%   ``Starburst'' & Geneva      & BaSeL 2.2 \\
%   ``BC03''      & Padova 1995 & STELIB \\
%   \enddata
%\end{deluxetable}

\clearpage

\begin{deluxetable}{ll}
   \tablecaption{Wavelength ranges that were not used for the spectral
   fit.\label{tab:cutouts}}
   \tablehead{range (\AA) & ``contamination'' source}
   \startdata
 3885-3900 & H8 \\
 3965-3980 & H$\epsilon$+[NeIII]3967 \\
 4100-4110 & H$\delta$ \\
 4335-4345 & H$\gamma$+[OIII]4363 \\
 4855-4870 & H$\beta$ \\
 4955-4965 & [OIII]4959 \\
 5000-5015 & [OIII]5007 \\
 5535-5590 & telluric lines \\
 5860-5905 & HeI 5876 \\
 6245-6320 & [OI]6300+[SIII]6312 \\
 6520-6600 & H$\alpha$+[NII] \\
 6700-6740 & [SII]6717+6731 \\
 7130-7145 & [ArIII]7136 \\
% vcc1175
% ArIII ? / SmI 7136.01 10 / PrII  7137.33 7 / ScI 7138.09 19 /
% NII 7138.85 - / TiI 7138.905 26 / CI 7139.18 - /
% AlII 7139.19 2 / SmII 7139.39 12 / SII 7139.787 12
 7560-7600 & telluric lines \\
 7700-8100 & telluric lines \\
 8250-8480 & telluric lines \\
\enddata
\end{deluxetable}

\begin{deluxetable}{ll}
  \tablecaption{Possible values of the population parameters.\label{tab:parameters}}
  \tablehead{parameter & possible values}
  \startdata
% $(M_{y}+M_{i})$:$M_{o}$    & 0:1, 1:100, 1:30, 1:10, 1:3, 1:1, 3:1, 10:1, 1:0 \\
 $(M_{y}+M_{i})$:$M_{o}$    & 0:1, 1:100, 1:30, 1:10, \\
                            & 1:3, 1:1, 3:1, 10:1, 1:0 \\
% $M_{y}$:$M_{i}$    & 0:1, 1:30, 1:10, 1:3, 1:1, 3:1, 10:1, 30:1, 1:0 \\
 $M_{y}$:$M_{i}$    & 0:1, 1:30, 1:10, 1:3, 1:1, \\
                    & 3:1, 10:1, 30:1, 1:0 \\
 $age_{y}$          & 1, 2, 3, 4, 5, 6, 7, 8, 9 Myr \\
 $age_{i}$          & 10, 20, 50, 100, 200, 500 Myr \\
 $age_{o}$          & fixed at 5 Gyr \\
% ${\rm [Fe/H]}_{y}={\rm [Fe/H]}_{i}={\rm [Fe/H]}_{o}$ & fixed at -0.3 \\
 ${\rm [Fe/H]}_{y}$ & fixed at -0.3 \\
 ${\rm [Fe/H]}_{i}$ & fixed at -0.3 \\
 ${\rm [Fe/H]}_{o}$ & fixed at -0.3 \\
  \enddata
  \tablecomments{
The young population is denoted by index $y$, the intermediate-age one by
 $i$ and the old one by $o$. $M_{x}$ is the mass fraction of the
 respective population.
  }
   \end{deluxetable}

\clearpage

   \begin{deluxetable}{lrrrrr}
     \tablecaption{Best fitting parameters using the ``BC99'' library.\label{tab:EWparametersbc99}}
     \tablehead{VCC & $M_{y}$ & $age_{y}$ & $M_{i}$ & $age_{i}$ & $M_{o}$}
     \startdata
\sidehead{Main sample}
 0021 & 0.0029 & 9 Myr & 0.0880 & 509 Myr & 0.9091 \\
 0281 & 0.0029 & 1 Myr & 0.0293 & 203 Myr & 0.9677 \\
 0781 & 0.0003 & 2 Myr & 0.0096 & 203 Myr & 0.9901 \\
 0870 & 0.0029 & 7 Myr & 0.0293 & 102 Myr & 0.9677 \\
 0951 & 0.0010 & 1 Myr & 0.0312 & 203 Myr & 0.9677 \\
 0953 & 0.0010 & 7 Myr & 0.0312 & 509 Myr & 0.9677 \\
 1078 & 0.0029 & 1 Myr & 0.0880 & 509 Myr & 0.9091 \\
 1488 & 0.0010 & 7 Myr & 0.0312 & 203 Myr & 0.9677 \\
 1684 & 0.0000 &   -   & 0.0909 & 203 Myr & 0.9091 \\
\sidehead{Additional sample}
%\colrule
%\tableline
 0024 & 0.0029 & 1 Myr & 0.0293 & 102 Myr & 0.9677 \\
 0135 & 0.0010 & 3 Myr & 0.0312 & 102 Myr & 0.9677 \\
 0340 & 0.0074 & 1 Myr & 0.0025 & 509 Myr & 0.9901 \\
 1175 & 0.0050 & 1 Myr & 0.0050 & 509 Myr & 0.9901 \\
 1273 & 0.0025 & 2 Myr & 0.0074 & 203 Myr & 0.9901 \\
 1437 & 0.0161 & 8 Myr & 0.0161 &  50 Myr & 0.9677 \\
 1499 & 0.0081 & 9 Myr & 0.2419 & 509 Myr & 0.7500 \\
 \enddata
  \tablecomments{
$M_{y}$ and $age_{y}$ give the mass fraction and age of the young
 population, respectively. $M_{i}$ and $age_{i}$ are the same
 parameters for the intermediate-age population. The age of the old
 population was fixed at 5 Gyr; $M_{o}$ gives its resulting mass
 fraction.
}
   \end{deluxetable}

%\clearpage

   \begin{deluxetable}{lrrrrr}
     \tablecaption{Best fitting parameters using the ``Starburst'' library.\label{tab:EWparameterssbmn}}
     \tablehead{VCC & $M_{y}$ & $age_{y}$ & $M_{i}$ & $age_{i}$ & $M_{o}$}
     \startdata
\sidehead{Main sample}
 0021 & 0.0083 & 8 Myr & 0.0826 & 500 Myr & 0.9091 \\
 0281 & 0.0029 & 1 Myr & 0.0293 & 200 Myr & 0.9677 \\
 0781 & 0.0003 & 8 Myr & 0.0096 & 100 Myr & 0.9901 \\
 0870 & 0.0227 & 8 Myr & 0.2273 & 500 Myr & 0.7500 \\
 0951 & 0.0029 & 9 Myr & 0.0880 & 500 Myr & 0.9091 \\
 0953 & 0.0025 & 8 Myr & 0.0074 & 100 Myr & 0.9901 \\
 1078 & 0.0083 & 8 Myr & 0.0826 & 200 Myr & 0.9091 \\
 1488 & 0.0029 & 8 Myr & 0.0880 & 500 Myr & 0.9091 \\
 1684 & 0.0029 & 1 Myr & 0.0880 & 500 Myr & 0.9091 \\
\sidehead{Additional sample}
%\colrule
%\tableline
 0024 & 0.0029 & 5 Myr & 0.0293 & 500 Myr & 0.9677 \\
 0135 & 0.0025 & 5 Myr & 0.0074 & 100 Myr & 0.9901 \\
 0340 & 0.0050 & 1 Myr & 0.0050 &  20 Myr & 0.9901 \\
 1175 & 0.0050 & 1 Myr & 0.0050 & 500 Myr & 0.9901 \\
 1273 & 0.0025 & 2 Myr & 0.0074 & 200 Myr & 0.9901 \\
 1437 & 0.0083 & 1 Myr & 0.0826 & 500 Myr & 0.9091 \\
 1499 & 0.0081 & 8 Myr & 0.2419 & 200 Myr & 0.7500 \\
 \enddata
  \tablecomments{
Same as Table~\ref{tab:EWparametersbc99}, but for the ``Starburst'' library.
}
   \end{deluxetable}

\clearpage

   \begin{deluxetable}{lrrrrr}
     \tablecaption{Best fitting parameters using the ``BC03'' library.\label{tab:EWparametersbc03}}
     \tablehead{VCC & $M_{y}$ & $age_{y}$ & $M_{i}$ & $age_{i}$ & $M_{o}$}
     \startdata
\sidehead{Main sample}
 0021 & 0.0029 & 9 Myr & 0.0880 & 509 Myr & 0.9091 \\
 0281 & 0.0029 & 1 Myr & 0.0880 & 509 Myr & 0.9091 \\
 0781 & 0.0003 & 1 Myr & 0.0096 & 203 Myr & 0.9901 \\
 0870 & 0.0029 & 6 Myr & 0.0293 & 203 Myr & 0.9677 \\
 0951 & 0.0010 & 6 Myr & 0.0312 & 203 Myr & 0.9677 \\
 0953 & 0.0010 & 7 Myr & 0.0312 & 509 Myr & 0.9677 \\
 1078 & 0.0029 & 1 Myr & 0.0880 & 509 Myr & 0.9091 \\
 1488 & 0.0010 & 7 Myr & 0.0312 & 203 Myr & 0.9677 \\
 1684 & 0.0029 & 6 Myr & 0.0880 & 509 Myr & 0.9091 \\
\sidehead{Additional sample}
%\colrule
%\tableline
 0024 & 0.0029 & 5 Myr & 0.0880 & 509 Myr & 0.9091 \\
 0135 & 0.0010 & 4 Myr & 0.0312 & 102 Myr & 0.9677 \\
 0340 & 0.0029 & 1 Myr & 0.0293 &  50 Myr & 0.9677 \\
 1175 & 0.0029 & 1 Myr & 0.0880 & 509 Myr & 0.9091 \\
 1273 & 0.0025 & 2 Myr & 0.0074 & 509 Myr & 0.9901 \\
 1437 & 0.0029 & 3 Myr & 0.0880 & 203 Myr & 0.9091 \\
 1499 & 0.0081 & 8 Myr & 0.2419 & 509 Myr & 0.7500 \\
 \enddata
  \tablecomments{
Same as Table~\ref{tab:EWparametersbc99}, but for the ``BC03'' library.
}
   \end{deluxetable}
%

%\clearpage

   \begin{deluxetable}{llllll}
\tablecaption{HI detections. \label{tab:HI}}
\tablehead{VCC  &  $log(M_{\rm HI}/M_{\odot})$ &
$\frac{M_{\rm HI}}{M_{\rm bary}}(\frac{M}{L}$=$3)$ &
$\frac{M_{\rm HI}}{M_{\rm bary}}(\frac{M}{L}$=$4.5)$ &
$\frac{M_{\rm HI}}{M_{\rm bary}}(\frac{M}{L}$=$6)$ &
Reference}
\startdata
\sidehead{Main sample}
0021  &  $<$7.78   &  $<$0.048 & $<$0.032 & $<$0.024 &   1 \\
0170  &  7.39      &  0.014    & 0.009    & 0.007 &   2\\
0281  &  7.52      &  0.055    & 0.037    & 0.028 &   2\\
      &  7.74      &           &          &       &   3\\
0308  &  $<$7.92   &  $<$0.094 & $<$0.065 & $<$0.049 &   1\\
0781  &  $<$7.42   &  $<$0.055 & $<$0.038 & $<$0.028 &   1\\
0951  &  $<$6.84   &  $<$0.003 & $<$0.002 & $<$0.001 &   1\\
0953  &  $<$6.54   &  $<$0.009 & $<$0.006 & $<$0.004 &   1\\
1488  &  $<$8.27   &  $<$0.134 & $<$0.093 & $<$0.072 &   1\\
1779  &  $<$7.72   &  $<$0.041 & $<$0.027 & $<$0.021 &   1\\ 
1912  &  $<$7.44   &  $<$0.031 & $<$0.021 & $<$0.016 &   1\\                
\sidehead{Additional sample}
%\colrule                                                            
0024  &  8.90      &  0.546 & 0.445 & 0.376 &   2\\
0135  &  $<$7.13   &  $<$0.009 & $<$0.006 & $<$0.004 &   2\\
      &  $\le$7.74 &        &       &       &   3\\
0334  &  7.89      &  0.339 & 0.255 & 0.204 &   2\\
0340  &  8.76      &  0.547 & 0.446 & 0.377 &   2\\
0446  &  7.62      &  0.284 & 0.209 & 0.166 &   2\\
0841  &  7.55      &  0.270 & 0.198 & 0.156 &   2\\
0890  &  7.27      &  0.106 & 0.073 & 0.056 &   2\\                          
1273  &  $<$7.10   &  $<$0.064 & $<$0.044 & $<$0.033 &   2\\
      &  $\le$7.54 &        &       &       &   3\\
1437  &  8.17      &  0.311 & 0.232 & 0.184 &   2\\
1499  &  $<$8.53   &  $<$0.256 & $<$0.186 & $<$0.147 &   4\\
1955  &  7.66      &  0.047 & 0.032 & 0.024 &   2\\
2007  &  7.31      &  0.218 & 0.157 & 0.122 &   2\\
      &  7.44      &        &       &       &   3\\
2033  &  7.39      &  0.045 & 0.031 & 0.023 &   2\\
\enddata
\tablecomments{
HI masses and upper limits are given for our adopted Virgo cluster
distance of $d=15.85 \rm{Mpc}$. In columns 3 to 5 we list the ratios of
the gas mass to the total baryonic mass, using different
mass-to-light ratios as given in the column header; for details see text.
References for the HI detections: 1. GOLDMine database \citep[][http://goldmine.mib.infn.it/]{goldmine};
2. \citet{gav05};
3. \citet{huc89};
4. \citet{huc86}.
%, $M_{HI}$ = 2.36 $10^{5}
%\times D^{2} \times$ S, where D is the distance assumed to be 21.8 Mpc
%and S is the HI flux integral in Jy $\times kms^{-1}$;
}
\end{deluxetable}

\clearpage

\end{document}